\documentclass[11pt,a4paper,oneside,notitlepage]{article}
\pdfoutput=1
\usepackage{jheppub}

\usepackage[colorlinks=true,citecolor=blue,linkcolor=blue]{hyperref}
\usepackage[T1]{fontenc}
\usepackage[normalem]{ulem}
\usepackage{amsmath,amssymb}
\usepackage{mathtools}
\usepackage{epsfig}
\usepackage{graphicx} 
\usepackage{grffile} 
\usepackage{url}
\usepackage{color}
\usepackage{multirow}
\usepackage{placeins}
\usepackage[dvipsnames]{xcolor}
\usepackage{epstopdf}
\usepackage{siunitx}
\usepackage{enumitem}
\usepackage{cleveref}
\usepackage{lipsum}
\usepackage{gensymb}
\usepackage{xspace}
\usepackage{titlesec} 
\usepackage{booktabs} 
\usepackage{subfigure}
\usepackage[subfigure,titles]{tocloft} 

\allowdisplaybreaks

\renewcommand{\vec}[1]{{\boldsymbol{#1}}}

\newcommand{\td}[2]{\frac{\mathrm d #1}{\mathrm d #2}}                


\begin{document}

\title{Probing Dark Sectors with Exploding Black Holes: Gamma Rays}

\author[a]{Michael J.~Baker,}   
\author[a]{Joaquim Iguaz Juan,}  
\author[a]{Aidan Symons,}
\author[a]{Andrea Thamm}     

\emailAdd{mjbaker@umass.edu}  
\emailAdd{jiguazjuan@umass.edu}
\emailAdd{asymons@umass.edu} 
\emailAdd{athamm@umass.edu}

\affiliation[a]{Department of Physics, University of Massachusetts Amherst, MA 01003, USA}

\date{\today}


\newpage

\abstract{
    The Hawking radiation from the explosion of a black hole would provide definitive information on the particle spectrum of nature.  Here we quantify the potential of current and future gamma ray telescopes to probe new dark sectors. We improve on the analysis used in previous work by making careful use of the experimental response functions, deriving a more realistic estimate of the backgrounds and optimizing the statistical analysis. We compute the sensitivity of the current experiments (HAWC and LHAASO) and estimate the reach of the future experiments (SWGO and CTA North and South), for various sky positions of the explosion.  We find that for a black hole exploding at $0.01\,\text{pc}$ the gamma ray signal observed by HAWC could probe dark sectors with 10-20 (or more) new Dirac fermions up to masses around $10^5\,\text{GeV}$, while CTA will be able to probe 2-15 new Dirac fermions with masses up to $10^6\,\text{GeV}$.  CTA North and South will have sensitivity to 10 dark fermions up to a distance of 0.1 \,pc and 50 up to a distance of 0.6\,pc.
}

\maketitle

\section{Introduction}
\label{sec:introduction}

Following the LIGO/Virgo~\cite{LIGOScientific:2016aoc} and EHT observations~\cite{EHT:2019dse} we now have direct evidence that black holes exist.  In 1974 Stephen Hawking showed that black holes are expected to emit particles with a quasi-blackbody distribution \cite{Hawking:1974rv,Hawking:1975vcx}. From black hole thermodynamics we know that the temperature of a black hole is inversely proportional to its mass.  Energy conservation means that as a black hole loses energy via Hawking radiation, it loses mass.  This increases its temperature, which increases the Hawking radiation emitted and leads to a runaway explosion process. 

All fundamental particles in nature with a mass roughly less than the black hole temperature will be emitted with a similar intensity (up to order one factors due to spin and internal degrees of freedom) irrespective of their non-gravitational couplings, so in the final moments of a black hole explosion it is expected that all particles present in nature (with masses below the Planck scale) will be radiated.  Even if these particles belong to a dark sector and do not produce Standard Model (SM) particles, they will affect the subsequent black hole evolution and leave an imprint on the gamma ray signal from the explosion~\cite{Ukwatta:2015iba,Baker:2021btk,Baker:2022rkn}. The Hawking radiation from an Exploding Black Hole (EBH) then, in principle, encodes information on all particles that exist in nature up to the Planck scale.

The solar mass and supermassive black holes that have already been seen~\cite{LIGOScientific:2016aoc,EHT:2019dse} will not explode any time soon, but a black hole produced in the early universe, a so-called primordial black hole (PBH), could be evaporating today.\footnote{Black hole lifetimes roughly scale as $\tau \propto M^3$.} There are many possible mechanisms of PBH formation~\cite{Hawking:1982ga, Carr:1975qj, Kawasaki:1997ju, Khlopov:2008qy, Brandenberger:2021zvn, Cotner:2018vug, Baker:2021nyl,
Baker:2021sno}, and PBHs have been invoked to solve many problems in cosmology and astrophysics (see, e.g., Refs.~\cite{Carr:2020gox,auffinger2022primordial,Matteri:2025vnv,Klipfel:2025jql,Baker:2025cff,Koulen:2025xjq,Yuan:2025avq,IguazJuan:2025vmd}). While indirect constraints on a population of $\sim 10^{15}\,\text{g}$ Schwarzschild black holes suggest that it may be optimistic to hope for an observation, these constraints rely on various assumptions.  In particular, it has been shown that if the black hole becomes quasi-extremal during its evolution before a final Schwarzschild-like burst, these constraints need not apply~\cite{Baker:2025zxm}. There have been many direct searches performed for exploding black holes, with a current best limit of 181 exploding black holes per cubic parsec per year in the vicinity of the Earth recently set by the LHAASO gamma ray telescope~\cite{Cao:2025oju}.  The previous best limit was $\sim 3400$\,pc$^{-3}$yr$^{-1}$ by the HAWC gamma ray telescope~\cite{HAWC:2019wla}.\footnote{ 
 This limit was derived using a simplified parametrisation of the total time integrated photon emission. A more careful analysis using \texttt{HDMSpectra} appears to strengthen the upper bound to $\sim\!1200$ bursts pc$^{-3}$yr$^{-1}$~\cite{Capanema:2021hnm}.} Over 3--10 years of observation, SWGO and CTA are projected to probe down to $\sim 50\,\text{pc}^{-3}\text{yr}^{-1}$ and $\sim 36\,\text{pc}^{-3}\text{yr}^{-1}$, respectively~\cite{Lopez-Coto:2021lxh,Yang:2025uvf}.  Given the pace of advance, we should be prepared for an observation of an exploding black hole in the near future.

Should the final moments of a black hole evaporation be observed by the current generation of high energy gamma ray telescopes, previous work has shown that with as few as $\sim 200$ photons they could exclude one new dark copy of the SM particles at a common mass scale up to $\sim 10^5\,\text{GeV}$~\cite{Baker:2021btk}.\footnote{This scenario has also been considered in Ref.~\cite{Federico:2024fyt}.}  In this work we extend this preliminary study by making a more realistic estimate of the backgrounds (predominantly from misidentified cosmic rays), improving the statistical analysis, considering a wide range of current and future gamma ray telescopes and exploring a range of sky positions of an explosion.  We do this for Beyond the Standard Model (BSM) scenarios containing a number of dark sector fermions at a common mass scale, which can be treated as proxies for the degrees of freedom of dark scalar or vector particles.\footnote{For a discussion of more realistic models beyond the SM see Ref.~\cite{Baker:2022rkn}.}  We find that the improvements in the gamma ray analysis substantially enhance the expected reach of HAWC, and that future observatories will significantly surpass HAWC's reach.

This paper is structured as follows: In \cref{sec:ParticleEmission} we review the theoretical expectations of Hawking radiation and the relevant computations for determining the flux of photons that would reach the Earth. In \cref{sec:Experiments} we review the essential features of a range of current and future gamma ray telescopes, and discuss their detection efficiencies and main backgrounds. \Cref{sec:probing-the-particle-spectrum} focusses on the signal and background rates. We compare and discuss four statistical analysis methods, and draw conclusions about the optimal photon binning. In \cref{sec:Results} we present our results for various sky positions of the explosion at HAWC, LHAASO, SWGO and CTA.

\section{The Photon Spectrum of an Exploding Black Hole}
\label{sec:ParticleEmission}

In this section we describe the photon spectrum expected from a Schwarzschild black hole of mass $M$, and discuss the evolution of the black hole mass with time assuming the SM or the SM plus a new dark sector. Black holes radiate all fundamental particles with a mass below their temperature.
As well as directly radiating primary photons, a black hole can emit other particles which can produce final state radiation, hadronise and/or decay to produce additional secondary photons. 
If there are new particles beyond the Standard Model they will also be radiated when the black hole temperature exceeds their mass, which will leave an imprint on the time evolution of the mass and temperature, and consequently on the emitted photons.

\begin{figure}
    \includegraphics[width = 0.49\textwidth]{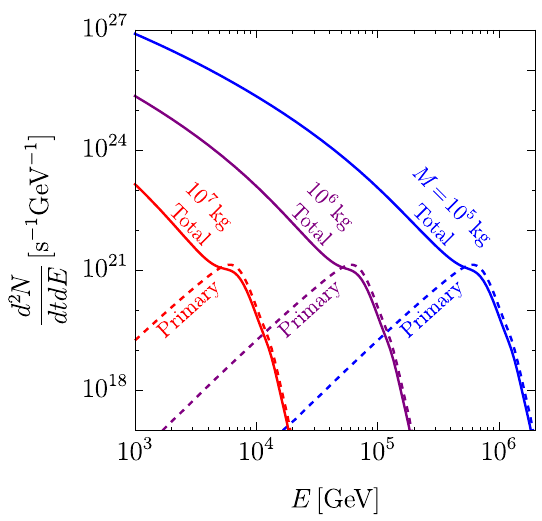}
    \hspace{0.02\textwidth}
    \includegraphics[width = 0.49\textwidth]{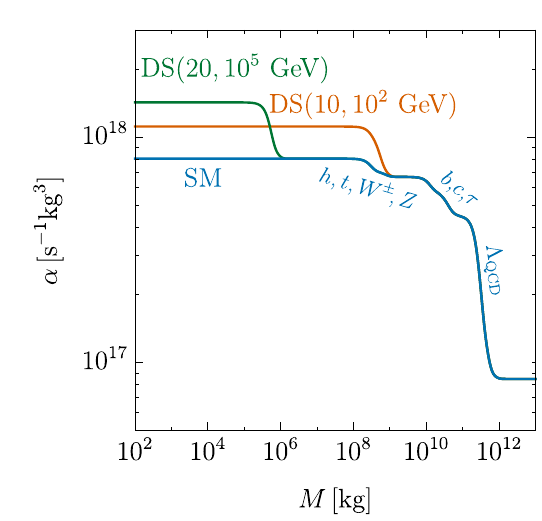}
    \caption{Left: The primary (dashed) and total (solid) gamma ray spectrum emitted from black holes of different masses.  Right: The Page function $\alpha(M)$, which encodes the particles present in nature, for the SM (blue) and with the SM plus two example dark sectors (orange and green), see text for details.}
    \label{fig:bh-emission}
\end{figure}

Since black holes typically radiate angular momentum and charge faster than mass~\cite{page:1976ratesII}, at the end of their life they are expected to be Schwarzschild.\footnote{Note that if there are many light or massless scalar fields present in nature,  Kerr black holes may lose mass at a higher rate than angular momentum, asymptotically approaching a non-zero value of angular momentum~\cite{TCH:1997ai,TCH:1998dk}.}  The mechanism discussed in Ref.~\cite{Baker:2025zxm}, where the indirect constraints on a population of PBHs do not apply, also features black holes that are Schwarzschild during their final explosion.  For these reasons, we will only consider the photons emitted by an exploding Schwarzschild black hole. These black holes radiate the fundamental particle species $i$ at a rate~\cite{Page:1976df}
\begin{align}
    \label{eq:primary-rate}
    \frac{d^2N_p^i}{dtdE}
    &=
    \frac{n_\text{dof}^i}{2\pi\hbar}
    \frac{\Gamma^i(M,E,s_i)}{e^{E/T}\pm1},
\end{align}
where $n_\text{dof}^i$ is the number of degrees of freedom of the radiated particle, $\Gamma^i(M,E,s_i)$ is the greybody factor which depends on the black hole mass $M$, the particle's energy $E$ and the particle's spin $s_i$, $T$ is the black hole's temperature, and there is a plus sign in the denominator for fermions and a minus sign for bosons. We are using units where $k_b=\hbar=c=1$. A Schwarzschild black hole has a temperature
\begin{align}
    T
    &=
    \frac{M_\text{Pl}^2}{8\pi M},
\end{align}
where $M_\text{Pl} = 1.218\times10^{19}\,\text{GeV}$ is the Planck mass.  The greybody factor $\Gamma^i(M,E,s_i)$ corresponds to the probability that a particle escapes from the black hole's event horizon to infinity.  Apart from a correction at $E \sim m_i$, where $m_i$ is the mass of the emitted particle, $\Gamma^i$ only depends on the ratio $E/T$.  Since particles with $E \sim m_i$ only make up a small proportion of the radiated particles, we neglect this effect.  We also set $\Gamma^i(M,E,s_i) = 0$ for $E < m_i$ (or $E < \Lambda_\text{QCD}$ for light coloured particles).  For $\Gamma^i(M,E,s_i)$ we take the values made publicly available in \texttt{BlackHawk}~\cite{Arbey:2019mbc,Arbey:2021mbl}.  \Cref{eq:primary-rate} then gives the spectrum of primary radiation emitted by a black hole of mass $M$, shown for photons in \cref{fig:bh-emission} (left) for several example black hole masses.
 
As the black hole emits Hawking radiation, conservation of energy means that it loses mass at a rate
\begin{align}
    \frac{dM}{dt}
    &=
    -\frac{\alpha(M)}{M^2},
\end{align}
where the Page function $\alpha(M)$ is given by
\begin{align}
    \alpha(M)
    &\equiv
    M^2
    \sum_i
    \int_{m_i}^\infty\,
    \!E\,
    dE
    \,\frac{d^2N_p^i}{dtdE}
    \,,
\end{align}
and where the sum is over all particles present in nature.  In this work we consider the SM particles (where we assume that the neutrinos are Majorana fermions), as well as a possible dark sector.  The dark sector does not necessarily contain a dark matter candidate, but acts as a stand-in for any new physics which does not couple significantly to the Standard Model particles. Since the particles couple to the black hole gravitationally, their emission does not depend on any gauge or Yukawa couplings between the dark sector and the SM particles.  For simplicity we imagine that the dark sector contains $n_\chi$ Dirac fermions at the common mass scale $\Lambda_\text{DS}$.  We then denote the dark sector $\text{DS}(n_\chi, \Lambda_\text{DS})$.  The Page function $\alpha(M)$ is shown for the SM and two dark sector models in \cref{fig:bh-emission} (right).

As well as primary photons, other particles emitted by the black hole will produce photons as they radiate, hadronise and/or decay.  The total rate of photon emission is given by
\begin{align}
    \frac{d^2N_{\text{total}}^\gamma}{dtdE}
    &=
    \mathcal{A}^{\gamma\to \gamma}(E)
    \frac{d^2N_p^\gamma}{dtdE}
    +
    \mathcal{A}^{{Z^0}\to \gamma}(E)
    \frac{d^2N_p^{Z^0}}{dtdE}
    +
    \sum_i
    \int_0^\infty
    \!dE'\,
    \frac{d^2N_p^i}{dtdE'}
    \frac{dN^{i\to \gamma}}{dE}(E,E')
    \,,
\end{align}
where the first term accounts for the loss of primary photons due to their final state radiation (so $\mathcal{A}^{\gamma\to \gamma}(E) < 1$ for all $E$), the second term accounts for $Z^0 \to \gamma$ conversion and the third term gives the secondary photons, which is computed by convolving the primary spectra of all species $i$ with their fragmentation functions $dN^{i\to \gamma}(E,E')/dE$.  We take the $\mathcal{A}(E)$ and fragmentation  functions from the public code \texttt{HDMSpectra}~\cite{Bauer:2020jay}, which is accurate in the energy range of interest ($E\gtrsim 100 \,\text{GeV}$).  The final photon spectra at Earth for several example black hole masses are shown in \cref{fig:bh-emission} (left).  We see that at high energies the total spectra are mostly due to the primary photons while at lower energies there is a large flux of particles produced by final state radiation, hadronisation and decay.  In this work we will assume that the dark sector particles in our BSM scenarios do not produce any photons after they are emitted, so they do not contribute to the secondary photon flux.

\section{Current and Future Experiments}
\label{sec:Experiments}

In this section we discuss a range of current and upcoming Cherenkov gamma ray observatories (HAWC, LHAASO, SWGO, CTA North and CTA South).\footnote{We do not include Pierre Auger as it will only detect a fraction of the gamma rays seen by the other experiments, since it is only sensitive to particles with energies above $\sim 10^8\,\text{GeV}$~\cite{Katharina_Holland_MSc_Thesis}.} HAWC and SWGO are primarily water based Cherenkov observatories, LHAASO is a hybrid water and electron-muon detector observatory while CTA North and South are atmospheric Cherenkov observatories.  In order to quantify to what extent an observation of an exploding black hole can probe BSM parameter space, we now discuss the relevant experimental specifications.

To have the highest chance of observing an exploding black hole, a gamma ray telescope needs to have a large effective area (to collect as many photons as possible), a large field of view (to observe a large portion of the sky at once) and a high duty cycle (so it is running as often as possible).
 To accurately measure the gamma ray signal, the telescope needs to have a large effective area, a good energy resolution (to accurately estimate the energy of the incoming gamma rays) and good timing resolution (to produce an accurate time series of the signal).   The background must also be as low as possible.  To reduce the background, telescopes need a good angular resolution (to reject background events coming from other regions of the sky) and a low cosmic-ray misidentification rate (since the main background to a gamma ray signal from an exploding black hole is misidentified cosmic rays~\cite{Abeysekara_2017}). 
 In \cref{tab:GammaRayExperiments} we give an overview of these specifications for the gamma ray observatories we consider.

We now discuss the relevant specifications of each observatory in detail.

\begin{table}
\footnotesize
\centering
\renewcommand{\arraystretch}{1.5} 
\setlength{\tabcolsep}{5pt} 
\begin{tabular}{ccccccc}
\hline\hline
 & HAWC & LHAASO & SWGO & \phantom{sp}& CTA North & CTA South \\
\hline \hline

Location [($\degree$N,$\degree$E)] & 
(19, -97) & 
(29, 100) & 
(-23, -68)&& 
(29, -18) & 
(-25, -70) \\

Completion [yr]& 
2015 & 
2021 &  
TBD && 
2018-2028& 
2025-2030 
\\

\hline

Effective Area @ $10^4$\,GeV [m$^2$] & 
$1\times10^4$ & 
$5\times10^5$ & 
$3\times10^5$ && 
$3\times10^6$ & 
$8\times10^5$ \\

Field of View [sr] & 
$\sim 3$ &
$\sim 3$ &
$\sim 3$ &&
$\sim 0.02\,(0.1)$ &
$\sim0.02\,(0.1)$ \\

Duty Cycle [\%]& 
$\sim100$ & 
$\sim100$ & 
$\sim100$ && 
$\sim15^\dagger$ & 
$\sim15^\dagger$ \\

Energy Resolution [\%]& 
$\lesssim 40$ & 
$\lesssim 40$ & 
$\lesssim 60$&& 
$\lesssim 10$ & 
$\lesssim 10$ \\

Time Resolution [ns]& 
$\lesssim 0.1$ & 
$\lesssim 0.5$ & 
$\lesssim 1$ && 
$\lesssim 1$ & 
$\lesssim 1$ \\

\hline

CR Effective Area at $10^4$\,GeV [m$^2$] & 
$2\times10^4$ & 
$5\times10^5$ & 
$1\times10^6$ && 
-- & 
-- \\

Angular Resolution [$\degree$]& 
$\lesssim 1$ & 
$\lesssim 1$ & 
$\lesssim 1$ && 
$\lesssim 0.1$ & 
$\lesssim 0.1$ \\

CR Misidentification Rate [\%]&
$\sim 0.2$ & 
$\lesssim 0.1$ &
$\lesssim 0.2$ &&
$\lesssim 1$ &
$\lesssim 1$ \\

\parbox{2.9cm}{\phantom{x}\\CR Background @ $10^4$\,GeV [GeV$^{-1}$s$^{-1}$]\\}&
$6\times10^{-10}$&
$1\times10^{-6}$&
$8\times10^{-8}$&&
$2\times10^{-10}$&
$8\times10^{-10}$\\
 
\hline

References & \cite{HAWC:2019xhp,Abeysekara_2017,HAWC:2022zma,HAWC:2024plu} 
& \cite{DiSciascio:2016rgi,LHAASO:2021xzn,Cao2021,Zhu:2023bex,LHAASO:2019qtb} 
&  
\cite{Albert:2019afb,Hinton:2021rvp,Mura:2023jse,SWGO:2025taj} 
&& \cite{Maier:2019afm,Gueta:2021vrf,Kobayashi:2021fed,Pecimotika:2022wpy,CTADates} 
& \cite{Maier:2019afm,Gueta:2021vrf,Kobayashi:2021fed,Pecimotika:2022wpy,CTADates}
\\

\hline \hline
\end{tabular}
    \caption{Approximate experimental specifications of the gamma ray observatories HAWC, LHAASO, SWGO, and CTA North and South. We give the cosmic ray backgrounds for the HAWC $\theta_1$ band and the sum of the SWGO zones.  A $^\dagger$ indicates our motivated assumptions.  See text for further details.
    }
    \label{tab:GammaRayExperiments}
\end{table}

\subsection{HAWC -- The High Altitude Water Cherenkov Observatory}

The High Altitude Water Cherenkov (HAWC) observatory is a gamma ray observatory located on the Sierra Negra, Mexico, consisting of an array of water Cherenkov detectors which observe high energy particle showers produced when highly energetic particles strike the atmosphere~\cite{HAWC:2013kzm}.  HAWC has been fully operational since 2015 and has searched for exploding primordial black holes in 959 days of data, setting an upper limit of $3400\,\text{pc}^{-3}\text{yr}^{-1}$~\cite{HAWC:2019wla}.

\begin{figure}
    \centering
    \includegraphics[width=0.7\textwidth]{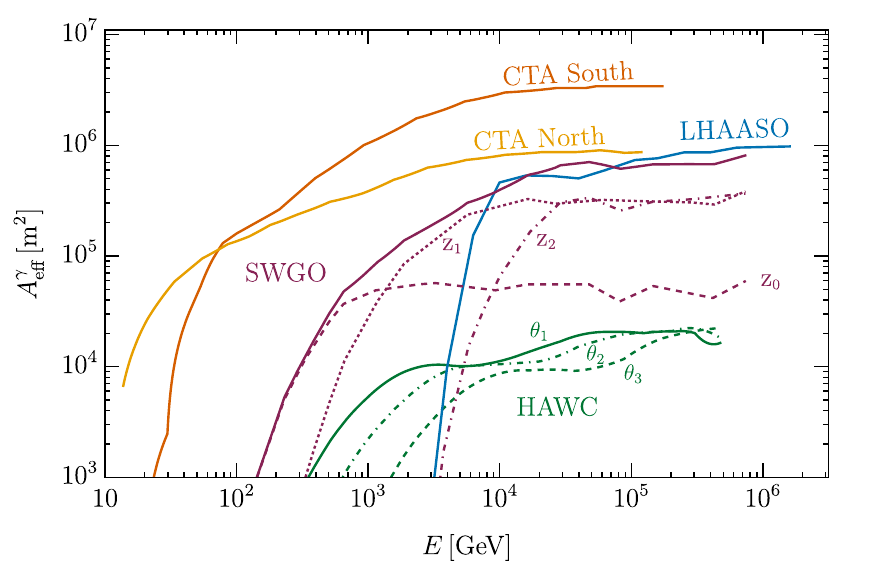}
    
    \caption{Effective areas for gamma rays versus the incoming gamma ray energy for the gamma ray telescopes HAWC, LHAASO, SWGO and CTA North and South.
    }
    \label{fig:effective-areas}
\end{figure}

In \cref{fig:effective-areas} we show HAWC's effective area for gamma rays as a function of the gamma ray energy~\cite{HAWC:2024plu}. Note that we use the on-array effective areas as these events are reconstructed with a better angular resolution.  The effective area depends on the incoming angle of the gamma ray (measured from the zenith of HAWC) and is given in three bands, labelled $\theta_1$ for $0 < \theta < 26\degree$, $\theta_2$ for $26\degree < \theta < 37\degree$, and $\theta_3$ for $37\degree < \theta < 46\degree$.  We see that HAWC has sensitivity to photon energies between 300\,GeV and $5\times10^5$\,GeV in the overhead $\theta_1$ band.  For gamma rays above $5\times10^5$\,GeV the photomultipliers in the detectors saturate and measurements can not be reliably made, so the effective area is set to zero.  HAWC has a large field of view of roughly 1.9\,sr, and is observing the sky almost all the time.  HAWC has fairly poor energy resolution with $\sim 20\%$ around $5\times 10^4\,$GeV, $\sim 40\%$ around $10^3\,$GeV and worse at lower energies~\cite{HAWC:2019xhp}, which is similar to other water-based Cherenkov observatories.  Because of this, we have designed analysis strategies that do not make use of the gamma ray energies.  
However, HAWC has excellent timing resolution, of the order of $100\,\text{ps}$~\cite{Marandon:2019sko}, which we use to bin events in short time intervals.

\begin{figure}
    \begin{tabular}{cc}
    \includegraphics[height=0.43\textwidth]{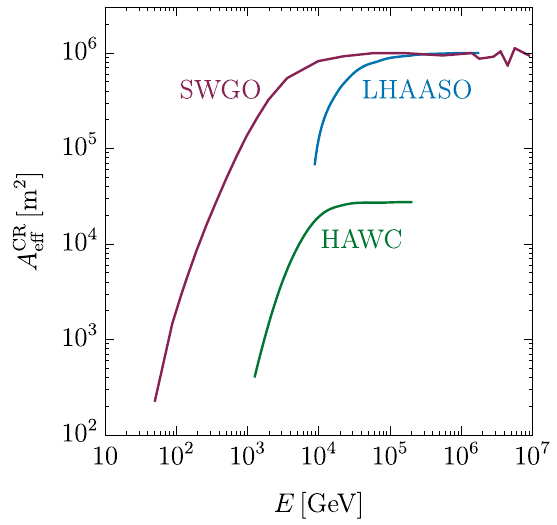}
    &
    \includegraphics[height=0.43\textwidth]{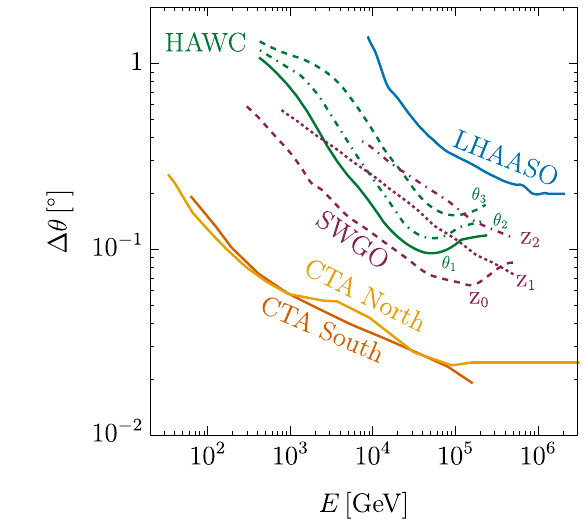}
    \\
    \includegraphics[height=0.43\textwidth]{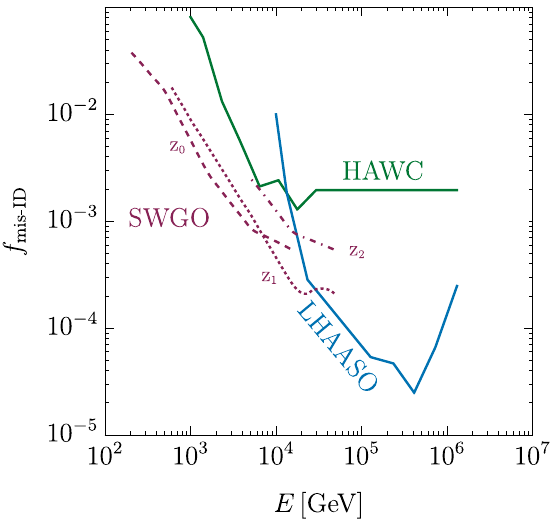} 
    &
    \includegraphics[height=0.43\textwidth]{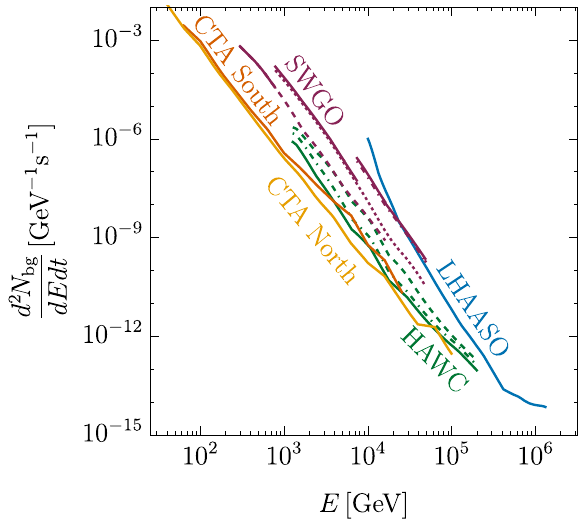}
    \end{tabular}
    \caption{Effective areas for light cosmic rays (top left), angular resolutions (top right), cosmic ray misidentification fractions (bottom left) and cosmic ray background rates (bottom right) versus the photon energy for the gamma ray telescopes HAWC, LHAASO, SWGO and CTA North and South (where available).
    }
    \label{fig:backgrounds}
\end{figure}

HAWC is able to distinguish heavy cosmic rays ($Z\geq3$) from photons and light cosmic rays. The effective area for light cosmic rays $A_\text{eff}^{\text{H}+\text{He}}$ is reported for HAWC in Ref.~\cite{HAWC:2022zma} and shown in the top left panel of \cref{fig:backgrounds} is the weighted average of the effective areas for H and He nuclei cosmic rays, averaged over zenith angles between $0\degree<\theta<16.7\degree$. Since the other theta bands are not given, we assume the effective area is independent of angle for $\theta\leq46\degree$.  We also ignore the unphysical dip at $E\gtrsim 10^{5.3}$ GeV seen in Ref.~\cite{HAWC:2022zma}, which is a simulation artefact.
HAWC's timing resolution helps it achieve a reasonably good angular resolution, $\Delta\theta(E) \lesssim 1\degree$, shown for the different $\theta$ bands in the top right panel of \cref{fig:backgrounds}.  We obtain these curves by interpolating the neural network containment radii given in Ref.~\cite{HAWC:2019xhp}. 
 To be conservative we take the worst angular resolution in a given energy range (the angular resolution also depends on the fraction of photomultiplier tubes that record light during the event).  HAWC also has a low cosmic ray misidentification rate of around $10^{-3}$~\cite{Abeysekara_2017}, as shown in the lower left panel of \cref{fig:backgrounds} (where the experimental paper does not distinguish the response in different $\theta$ bands). 
The HAWC experiment presents the cosmic ray misidentification rate as a function of the fraction of photomultiplier tubes that record light during the event, $\mathcal{B}$, instead of the energy of the incoming gamma rays.  Whilst experimental analyses simulate incoming gamma rays and cosmic rays in their detector, we approximate the angular resolution and cosmic ray misidentification fraction by fitting the peaks of the energy distributions in fig.~2 of Ref.~\cite{Abeysekara_2017}, giving
\begin{equation}
    \mathcal{B}(E) = a\log_{10}\left(\frac{E}{\text{GeV}}\right)+b,
    \qquad a=1.97175\,,
    \,\, 
    b=-11.2911
    \,.
\end{equation}
This allows us to approximately convert from the specifications given as a function of $\mathcal{B}$ to a function of the true incoming energy, $E$.  The curves shown in the lower left panel of \cref{fig:backgrounds} agree fairly well with the specifications HAWC estimated before running their experiment~\cite{SensitivityHAWC}.

The main background to the gamma ray signal from an exploding black hole is due to misidentified hadronic cosmic rays since their flux is much larger than the background gamma ray or electron/positron fluxes.  The main hadronic cosmic ray backgrounds are hydrogen and helium nuclei since heavier nuclei are easier to distinguish from gamma rays.  We estimate the background rate of these misidentified cosmic rays within the angular resolution of the detector over the duration of observation using~\cite{Yang:2024vij},
\begin{equation}
\label{eq:cr-background}
    \frac{d^2N_\text{bg}}{dEdt}
    =
    \td{\Phi(E)}{E}\,
    A_\text{eff}^{(\text{H}+\text{He})}(E)\,
    2\pi\big(1-\cos{(\Delta \theta(E)}\big)\,
    f_\text{mis-ID}(E),    
\end{equation}
where $d\Phi(E)/dE$ is the differential diffuse flux of light cosmic rays, $A_\text{eff}^{(\text{H}+\text{He})}(E)$ is the effective area for the light cosmic rays (which for simplicity we take for the zenith band $0 < \theta < 16.7\degree$), $\Delta \theta(E)$ is the angular resolution, and $f_\text{mis-ID}(E)$ is the fraction of incoming cosmic rays misidentified as photons. 
For the diffuse cosmic ray flux we use a parametrisation of the isotropic background~\cite{HAWC:2022zma},
\begin{equation}
\label{eq:cr-flux}
    \frac{d\Phi(E)}{dE}
    =
    \Phi_0E^{\gamma_1}
    \left[
    1 + 
    \left(\frac{E}{E_0}\right)^{\epsilon}
    \right]^{\frac{\gamma_2-\gamma_1}{\epsilon}},
\end{equation}
where 
\begin{equation}
\begin{aligned}
    \Phi_0 &= 10^{3.71}\text{\,m}^{-2}\text{\,s}^{-1}\text{\,sr}^{-1}\text{\,GeV}^{-1}\,, \qquad  E_0 = 10^{4.38}\text{ GeV} \,,\\ 
    \gamma_1 &= -2.51\,, \qquad
    \gamma_2 = -2.83\,, \qquad
    \epsilon = 9.8
    \,.
\end{aligned}
\end{equation}
This flux is fit to data up to $E \sim 10^5$\,GeV, but we use the same parametrisation above this energy scale. We will see that there is only a small contribution above this scale. We assume that the cosmic ray flux is isotropic, since the trajectories of charged particles are bent by interstellar magnetic fields with radii shorter than galactic length scales.  The differential background rate is then shown in the lower right panel of \cref{fig:backgrounds}.  We see that the background rate falls sharply as a function of energy, primarily due to the reducing cosmic ray flux.

\subsection{LHAASO -- The Large High Altitude Air Shower Observatory}

The Large High Altitude Air Shower Observatory (LHAASO) is a collection of cosmic ray and gamma ray detector arrays designed for making ultra-high energy observations in the PeV range \cite{DiSciascio:2016rgi}. Located $4410\,\text{m}$ above sea level in the Sichuan province of China, LHAASO has been taking measurements since 2019 and has so far detected and identified likely sources of over a dozen $\mathcal{O}(1\,\text{PeV})$ photons \cite{LHAASO:2021xzn,Cao2021}. LHAASO consists of two subarrays: the Kilometer-Square Array (KM2A) and the Water Cherenkov Detector Array (WCDA) composed of electron and muon detectors. In the following, we consider only KM2A as an independent detector since the cosmic ray misidentification fraction, $f_\text{mis-ID}$, is unavailable for WCDA which is crucial for a reliable background estimate. 

The LHAASO effective area for gamma rays is shown in \cref{fig:effective-areas}. For the KM2A subarray, it is a factor of around 50 larger than at HAWC for energies above $10^4$\,GeV.  We also see that LHAASO can detect gamma rays up to energies around $10^6$\,GeV. 
 Unlike HAWC, LHAASO does not provide enough information to compute the signal and background in different zenith bands so we only consider the central region.  Like HAWC, LHAASO has a large field of view and a high duty cycle.  The energy resolution of the KM2A subarray of LHAASO is better than 40\% at all energies above 10\,TeV~\cite{LHAASO:2019qtb}.  Like HAWC, LHAASO has sub-nanosecond timing resolution.

In the top left panel of \cref{fig:backgrounds} we show the effective area for cosmic rays of the  KM2A subarray of LHAASO.   To account for heavier hadronic cosmic rays (Helium and heavier elements) we follow \cite{Yang:2024vij,Abdo:2014apa} and multiply the background rate due to protons in \cref{eq:cr-background} by $1.2$.  We see that it is over an order of magnitude larger than HAWC.
In the top right panel of \cref{fig:backgrounds} we show the angular resolution.  We see that the KM2A resolution is somewhat worse than HAWC.  In the bottom left panel we show the LHAASO KM2A cosmic ray misidentification fraction and see that thanks to additional muon detectors its misidentification faction is more than an order of magnitude better than HAWC's above $\sim 10^4$\,GeV.  
Finally in the bottom right panel we show the background rate at LHAASO, obtained using \cref{eq:cr-background,eq:cr-flux}.  We see that the larger cosmic ray effective area is only somewhat counteracted by the improved cosmic ray misidentification fraction, leading to a background rate that is one to three orders of magnitude larger than HAWC's.

\subsection{SWGO -- The Southern Wide-field Gamma-ray Observatory}
\label{sec:SWGO_Overview}

The Southern Wide-field Gamma-ray Observatory (SWGO) is a planned water Cherenkov gamma ray telescope.  It aims to cover the 100\,GeV to PeV scale and observe a significant fraction of the southern sky with a wide field of view and high duty cycle~\cite{SWGO:2025taj}. It will be located in northern Chile at an altitude of 4770\,m.  While the completion date has not yet been determined, seven of the nine R\&D milestones have been completed~\cite{SWGO:2025taj}.

SWGO will consist of three different zones of water Cherenkov detectors.  The central zone, $z_0$, will have the highest density of detectors, and the density reduces in the outer two zones, $z_1$ and $z_2$. The effective areas for gamma rays of the three zones are shown in \cref{fig:effective-areas}.  When combined (shown in solid purple), the three zones have an effective area approximately the same as LHAASO above $10^4$\,GeV and a larger effective area than LHAASO below this.  SWGO's energy resolution is mildly better than HAWC's at higher energies, with $\sim 15\%$ around $5\times 10^4\,$GeV and $\sim 60\%$ around $10^3\,$GeV~\cite{SWGO:2025taj}. 
The array will also achieve nanosecond timing resolution~\cite{Mura:2023jse}.

In the top left panel of \cref{fig:backgrounds} we show the effective area for proton cosmic rays for SWGO as a whole.  We see that this area is very similar to that of LHAASO.  We again account for heavier hadronic cosmic rays by multiplying by $1.2$.  SWGO's angular resolutions are shown in the top right panel for zenith angles less than $30\degree$.  We see that the central zone will have slightly better angular resolution than HAWC.  In the bottom left panel we show the cosmic ray misidentification fractions of the three zones, which improve on HAWC's rejection efficiency.  These fractions are only available up to around $10^4$--$10^5$\,GeV due to their limited Monte Carlo statistics but, given the strongly falling astrophysical flux, cosmic rays above this energy make a negligible contribution.  We show the background rates, computed using \cref{eq:cr-background,eq:cr-flux}, in the lower right panel.  We see that due to the large effective area, the background rate will be around an order of magnitude larger than HAWC's over the energy range where they are both sensitive.

\subsection{CTA -- The Cherenkov Telescope Array}
\label{sec:CTA_Overview}

The Cherenkov Telescope Array (CTA) is an upcoming Imaging Atmospheric Cherenkov Telescope (IACT) situated at two sites: CTA North on the island of La Palma, Spain, and CTA South near Cerro Paranal, Chile \cite{Gueta:2021vrf,Hofmann:2023fsn}. The array will consist of over 100 Cherenkov telescopes of various sizes, each optimised for different gamma ray energies. Both sites will be sensitive to photons with energies between roughly $20$ and $3\times10^5\,\text{GeV}$ and will operate in two distinct modes: co-pointing and diverging.  In the co-pointing mode all telescopes focus on one point in the sky. While this has good performance in terms of, e.g., effective area, energy resolution and angular resolution, it has a small field of view of around 0.02\,sr \cite{Gueta:2021vrf}. In the diverging mode the component telescopes focus on different patches of the sky and thereby increase the combined field of view to around 0.1\,sr while sacrificing performance \cite{CTAConsortium:2023ybn}.  It is not yet known how much time CTA will spend in the two configurations.  In this work we will assume that CTA is operating in the co-pointing mode, and that both North and South arrays can be directed to focus upon any point in the sky up to $60\deg$ from their respective zeniths.  We note that although CTA North and South have small fields of view in co-pointing mode, their large effective areas and low background rate mean they are sensitive to exploding black holes in a similar volume of space to HAWC and LHAASO.  Depending on the energy range considered, the search volumes of HAWC, LHAASO and CTA North and South are all around $10^{-3}\,\text{pc}^3$ while SWGO's is around an order of magnitude larger, $10^{-2}\,\text{pc}^3$.  This means that each CTA site is as likely as HAWC to see an exploding black hole by chance.  While in this work we consider the observatories in isolation, in the future it would be of interest to analyse synergies between these (and other) experiments.

We show the gamma ray effective area of CTA North and South in co-pointing mode in \cref{fig:effective-areas}.  We see both CTA sites are focussed on lower energy photons than the water Cherenkov observatories, but still have very large effective areas in the $10^4$ -- $2\times 10^5$\,GeV energy range.  As discussed above both CTA sites have a field of view of around 0.02\,sr in co-pointing mode.  Although their duty cycles are not yet known, we can assume that they will be similar to other IACT's like H.E.S.S., which is around 15\%~\cite{Ohm:2023cgv}.  The energy resolution of both sites is significantly better than at the water Cherenkov detectors, attaining a $\sim$5\% energy resolution at $10^3$\,GeV~\cite{Gueta:2021vrf}.  We assume CTA will also have nanosecond timing resolution.

In the top right panel of \cref{fig:backgrounds} we show the angular resolution of both CTA sites.  We see that it surpasses the water Cherenkov detectors by almost an order of magnitude.  While CTA has not published the cosmic ray effective areas and cosmic ray misidentification rates, they do publish their cosmic ray background rate \cite{Maier:2019afm}, shown in the bottom right panel in \cref{fig:backgrounds}.  
Note that while the background at water Cherenkov detectors is dominated by hadronic cosmic rays, the background for CTA largely consists of $e^\pm$ cosmic rays between $200 \text{ GeV}\lesssim E \lesssim 1.5\times10^3$ GeV. 
We see that once all these factors are taken into account, both CTA sites have a very similar background rate to HAWC.  Note that this background rate is for the diffuse CR flux. CTA is expected to spend much of its time in co-pointing mode observing luminous objects of interest, which may impact the background rate. However, if particles from such a source come from a small enough region of the field of view, then our estimate for the search volume should not be severely impacted. For simplicity we ignore this complication.

\subsection{Sky Map}

\begin{figure}
    \centering
    \begin{tabular}{cc}
    \includegraphics[width=0.975\textwidth]{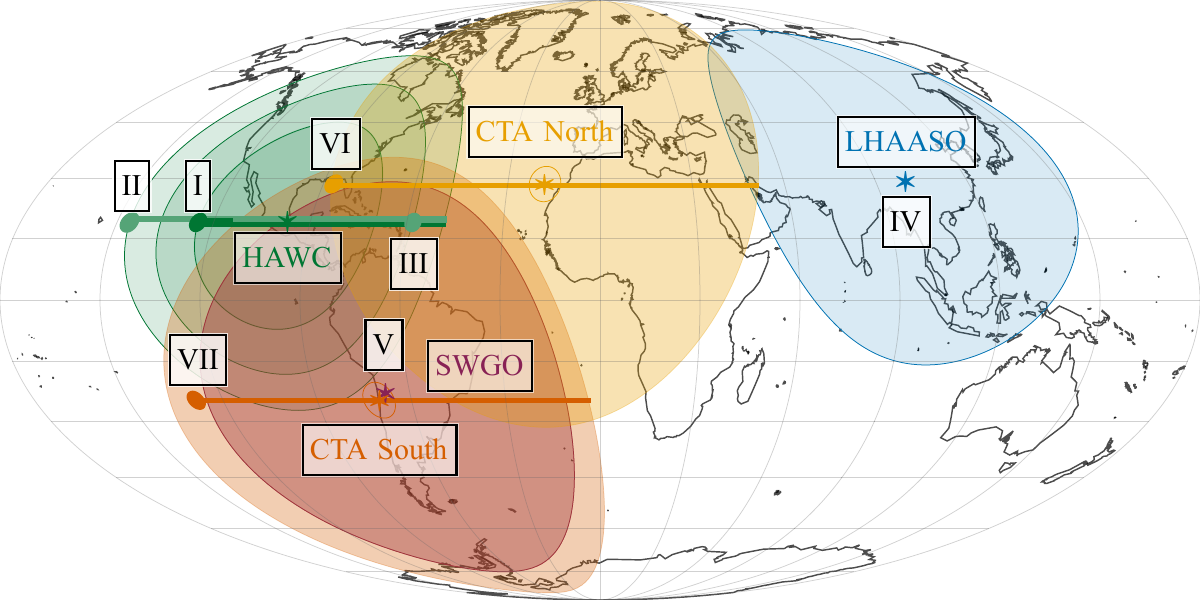}\\
    \end{tabular}
    \caption{
    Map of the sky (presented as the point in the sky directly above the corresponding point on the surface of the Earth) with current and projected fields of view as given in \cref{tab:GammaRayExperiments}. See text for details.}
    \label{fig:SkyMap}
\end{figure}

In \cref{fig:SkyMap} we show the regions of sky each observatory is sensitive to using the equal-area Mollweide projection. Since the observatories are all terrestrial, we show the outlines of the land masses on Earth, and the regions of sensitivity are the regions of sky directly above the regions shown on the surface of the Earth. The green, blue and red regions show the fields of view of the water Cherenkov observatories HAWC, LHAASO and SWGO, respectively. The yellow and orange regions show the regions that CTA North and South could point to while in co-pointing mode. We show a solid yellow and orange circle within this to illustrate CTA's field of view at any one time.  We see that HAWC and SWGO have a significant overlap with both CTA North and South while LHAASO observes a complementary region of the sky.  In particular, SWGO and CTA South can view almost the same patch of sky, so there is potential for a strong synergy between these experiments.

In this work we will imagine that a black hole explodes in one of seven different sky locations.  We will see that the background at each experiment will determine how long the explosion can be observed for at each experiment.  We indicate with lines how the exploding black hole will move in the sky during the observation window we determine below. The final explosion is in HAWC's main field of view in scenario I.  Scenarios II and III are less favourable for HAWC as the explosion occurs just before the black hole leaves HAWC's field of view and just before it enters the $\theta_2$ band, respectively. 
Scenarios IV and V, indicated by a star, refer to explosions directly above the experimental sites of LHAASO and SWGO, respectively. As discussed below, due to a short integration time in LHAASO and SWGO there are no associated lines.
Finally, in scenarios VI and VII we will consider the sensitivity of CTA North and CTA South.

The time an exploding black hole spends in a given field of view region can be determined by calculating the difference in longitude between the points where the field of view intercepts the black hole's trajectory (or point at which the black hole totally evaporates, if this occurs inside the field of view region under consideration). We provide details of this geometric calculation in \cref{sec:AppendixFOVGeometry}.

\section{Probing the Particle Spectrum with an Exploding Black Hole}
\label{sec:probing-the-particle-spectrum}

In this section we will discuss how these observatories could use an observation of an explo-ding black hole to probe the presence of a dark sector.  We will assume that the experiment has already determined that an observation is most likely the result of a black hole explosion.  This can be done by observing a hardening of the photon spectrum with time (which is not typical from known astrophysical sources), observing no afterglow with follow-up observations (which is typical of known astrophysical sources) and being unable to associate the burst with any known source.  Here we will discuss the signal from an exploding black hole in different BSM scenarios, how the background rates at each observatory impact the observation time, the most reliable statistical analysis and how the signal could be analysed to best probe the dark sector.

\subsection{Signal from an Exploding Black Hole}
\label{sec:EBHSignalPhotons}

Since the start of the explosion is gradual, it makes sense to measure time from the end of the explosion, which is very sudden.  We therefore measure time using $\tau = t_\text{end} - t$, where $t_\text{end}$ is the time of the end of the explosion so that the explosion occurs at $\tau = 0$\,s and larger $\tau$ correspond to earlier times.  The total number of photons observed from an exploding black hole by a given telescope, $N_\gamma$, in the energy window $(E_\text{min},E_\text{max})$ and time window $(\tau_\text{min},\tau_\text{max})$, is given by
\begin{equation}
    N_\gamma
    =
    \frac{1}{4\pi d^2}
    \int_{E_{\text{min}}}^{E_{\text{max}}}
    \!dE\,
    A_{\text{eff}}^\gamma(E,\theta)
    \int_{\tau_{\text{min}}}^{\tau_{\text{max}}}
    \!d\tau
    \frac{d^2 N_\text{total}^\gamma}{d\tau dE}
    \,,
    \label{eq:EBHSignal}
\end{equation}
where $d$ is the distance of the black hole from Earth, $A_\text{eff}^\gamma(E,\theta)$ is the energy and angle dependent photon effective area of the given telescope (see \cref{fig:effective-areas}), and $d^2 N_\text{total}^\gamma/dt dE$ is the total emission rate of photons. Of these quantities, $d$ depends on the distance to the explosion, which cannot be controlled, $A_{\text{eff}}^\gamma(E,\theta)$ depends on the design of the observatory and $d^2 N_\text{total}^\gamma/dt dE$ depends on the particles present in nature. We assume that the photon spectrum does not change en route to Earth, which is a good assumption for the distances we are considering.  Although the energy resolution of the telescopes we consider is at the $10-50$\% level, as a first approximation we do not consider energy smearing.  We now consider the appropriate choices for $E_\text{min},E_\text{max},\tau_\text{min}$ and $\tau_\text{max}$.

\subsection{Background Rates} 
\label{sec:BGRates}

In this work we conservatively choose the integration limits for the time and energy integration so that the signal is in a background free regime.  That is, the total integration time is chosen so that the expected number of background events is less than one.

\begin{figure}
    \centering
    \includegraphics[width=0.475\textwidth]{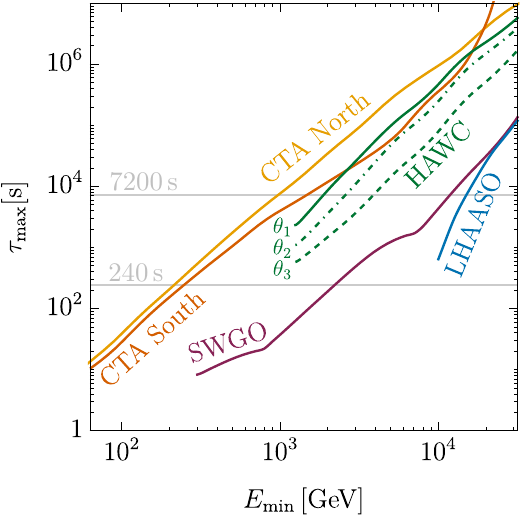}
    \caption{Maximal integration times versus the minimum photon energy for the gamma ray telescopes HAWC, LHAASO, SWGO, and CTA North and South. The horizontal gray lines at $\tau_\text{max} = 240\,$s and $\tau_\text{max} = 7200\,$s indicate the time in which the Earth rotates by more than $1\degree$ (the approximate angular resolution of the experiments) and by one grid line of longitude in \cref{fig:SkyMap}, respectively.
    }
    \label{fig:MaxIntegrationTime}
\end{figure}

Assuming that the cosmic ray background is time independent, the number of expected background events observed is approximately proportional to the duration of observation $\tau_\text{max}-\tau_\text{min}\approx\tau_\text{max}$ (since $\tau_\text{max} \gg \tau_\text{min}$),
\begin{align}
\label{eq:background}
    N_\text{bg}^\gamma
    &= \int_{\tau_{\text{min}}}^{\tau_{\text{max}}} dt
    \int_{E_{\text{min}}}^{E_{\text{max}}} 
    dE\,\frac{d^2N_\text{bg}}{dEd\tau}
    \approx 
    \tau_\text{max}
    \int_{E_{\text{min}}}^{E_{\text{max}}} 
    dE\,\frac{d^2N_\text{bg}}{dEd\tau}
    \,,
\end{align}
where $d^2N_\text{bg}/dE dt$ is given by \cref{eq:cr-background}.  The lower time limit $\tau_\text{min}$ can, in principle, be taken as the time resolution of the telescope. However, in realistic scenarios very few photons are observed for $\tau\lesssim10^{-6}\,\text{s}$, and so we take $\tau_\text{min}=10^{-6}\,\text{s}$.  We choose the upper energy limit $E_\text{max}$ to be the maximum energy for which we can compute the background rate for each observatory, see \cref{fig:backgrounds}.  Since the backgrounds fall quickly with energy the backgrounds become approximately independent of $E_\text{max}$ at these values as long as $E_\text{min} \ll E_\text{max}$.

For a given $E_\text{min}$ we can then set $N_\text{bg}^\gamma = 1$ and use \cref{eq:background} to determine $\tau_\text{max}(E_\text{min})$, which we show in \cref{fig:MaxIntegrationTime}.   We see that as $E_\text{min}$ is increased, the maximum integration time increases.  HAWC and CTA North and South have the longest integration times, since they have the lowest background rates.  The larger backgrounds of LHAASO and SWGO mean that they have shorter integration times.  We show with horizontal lines where the black hole will move by one degree (the approximate angular resolution of the water Cherenkov experiments) and by one grid line of longitude in \cref{fig:SkyMap} due to the rotation of the Earth.  We see that HAWC and CTA North and South expect to see the black hole move a large distance across the sky, while for SWGO and LHAASO this depends on $E_\text{min}$.  In the next section we will look at the impact of varying $E_\text{min}$ (and consequently $\tau_\text{max}$).

\subsection{Analysis Strategies} 
\label{sec:analysis-strategies}

In this section we discuss the details of our statistical analysis. We discuss a range of statistical tests and explore the sensitivity achieved through different binning schemes.

If an exploding black hole is observed, an experiment will observe a burst of gamma rays from a fixed point in the sky (with respect to the background stars).  If we assume the black hole is Schwarzschild and only the SM particles exist, then all the quantities in \cref{eq:EBHSignal} are known except for $d$ and we can use the number of photons observed and known details of the observatory to infer the distance to the black hole.  However, if we want to test various hypotheses of particles beyond the SM, then $N_\gamma$ suffers from a degeneracy: there could be fewer photons observed than expected either because BSM degrees of freedom are changing the black hole's emission, or simply because it is further away.  The total number of photons observed cannot be used to differentiate the SM from BSM hypotheses, but the signal must be binned and the shape must be used to differentiate between them.  We have the option of binning in time and/or energy, but since the energy resolution of the water Cherenkov telescopes is not very good, in this work we choose to only bin in time.  Furthermore, since the photons produced by final state radiation, hadronisation and/or decay dominate over primary photons for $E\lesssim T_{BH}$, it is difficult to extract useful information from the energy spectrum.  However, it would be interesting to investigate the extra sensitivity that could be gained by using the energy information, especially at IACT's such as CTA North and South.

Due to the degeneracy discussed above, we assume one of the experiments observes a fixed number of photons.  We parameterise this number of photons by the distance to the black hole assuming the SM hypothesis, $d_\text{SM}$.  We then assume other observatories observe the number of photons corresponding to an explosion at this same distance, $d_\text{SM}$ (again under the SM hypothesis).  This lets us compare different observatories, which have, e.g., different effective areas and so will observe a different number of photons from the same event, on a like-for-like basis.  

\subsubsection{Comparison of Different Statistical Tests}

\begin{figure}
    \centering
    \includegraphics[width=0.5\textwidth]{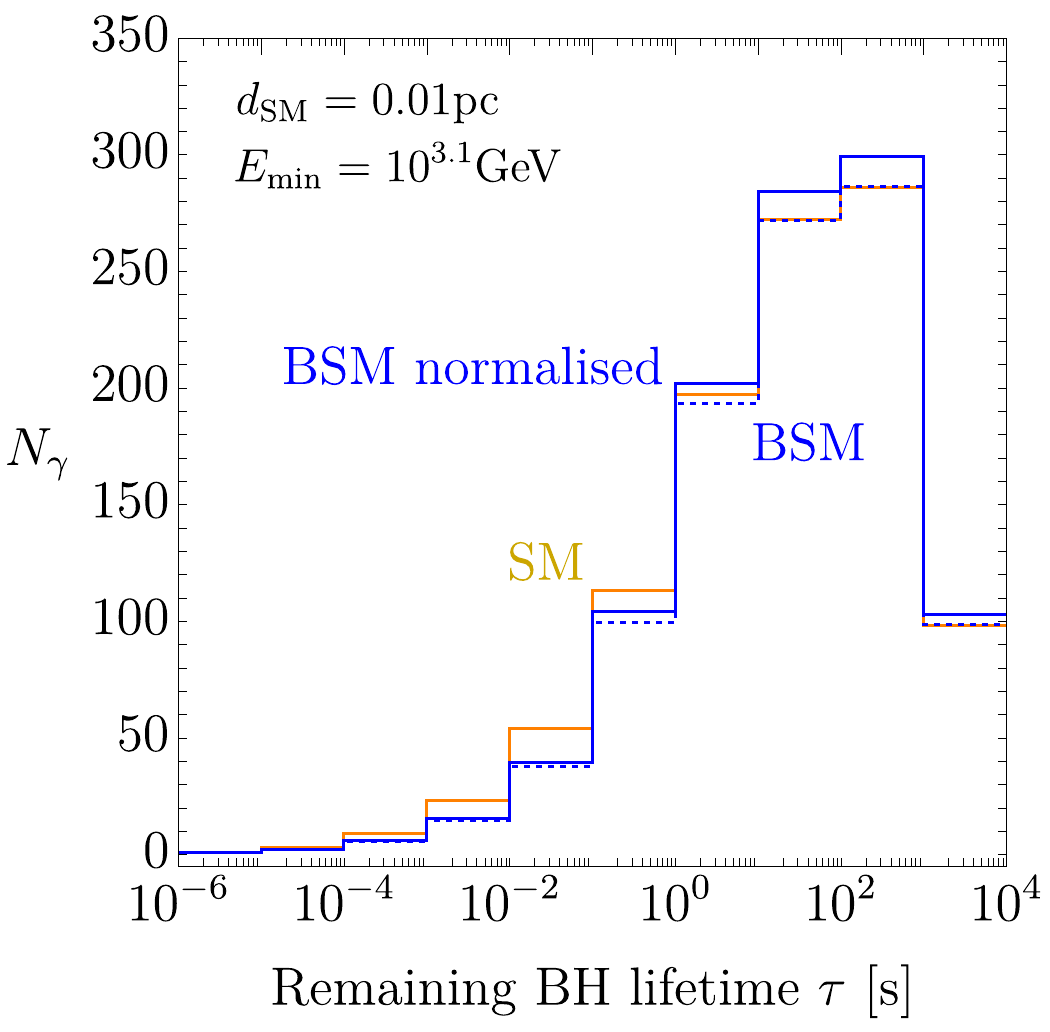}
    
    \caption{The signal from an exploding black hole seen in the central portion of HAWC with decadal binning, at a distance of $d_\text{SM} = 0.01$\,pc in the SM scenario (orange) and $d_\text{DS} = 0.0098$\,pc in the DS(30, $10^5$\,GeV) scenario (solid blue).  We also show the signal in the DS(30, $10^5$\,GeV) scenario at a distance of $d_\text{DS} = 0.01$\,pc (dashed blue).
    }
    \label{fig:decadal-binning}
\end{figure}

Previous work in this area \cite{Baker:2021btk} has used decadal binning, where photons are grouped into bins with edges at $\{10^{-6}$\,s$,10^{-5}$\,s$,10^{-4}$\,s$,...,\tau_{max}(E_\text{min}) \}$.  We show the number of photons with decadal binning for HAWC in \cref{fig:decadal-binning}, where we assume $d_\text{SM} = 0.01$\,pc and we take $E_\text{min} = 10^{3.1}$\,GeV (where $\tau_\text{max} = 2346$\,s).  We show both the SM signal (orange) and a BSM signal (blue), assuming a dark sector with 30 dark Dirac fermions at a mass scale of $10^{5}$\,GeV, DS(30, $10^5$\,GeV). In the SM the total number of photons observed is 1056.
In dashed blue we show the BSM signal  at $d_\text{DS} = 0.01$\,pc  which corresponds to 1010 total photons. As mentioned above, we see that the presence of the dark sector reduces the total number of photons observed (since the new degrees of freedom increase $\alpha$, which speeds up the explosion, so the black hole spends a shorter time at higher temperatures).  Since the black hole starts emitting BSM particles of mass $10^5$\,GeV at $\tau\approx 10$\,s~\cite{Baker:2022rkn}, this effect can be observed for the bins below $10$\,s.  We also show the BSM signal when the total number of photons is set equal to the SM signal of 1056 (solid blue), corresponding to an explosion at $d_\text{DS} = 0.0098$\,pc under the BSM scenario.  
This normalisation leads to an excess of photons at $\tau > 1$\,s and a deficit at $\tau < 1$\,s compared to the SM.

We also see in \cref{fig:decadal-binning} that there are hundreds of photons in each bin for $\tau \gtrsim 10^{-1}$\,s, but that the number drops quickly at times closer to the end of the explosion.  This means that some bins contain very few photons, where we cannot assume that statistical fluctuations are Gaussian distributed.  We now turn to statistical tests of the two distributions, where this point becomes relevant.

Without any actual data, we assume that the observed photon signal agrees with the Standard Model prediction.  We can then use a goodness of fit test to quantify the probability 
that the SM signal would be observed when the BSM model is the true model of nature.  The most common of these tests is the $\chi^2$-test, which works well when the statistical errors are Gaussian distributed.  Due to the central limit theorem, this typically applies when there are more than tens of events in each bin.  In our case, however, decadal binning leads to bins with fewer than ten events.  We therefore compare several more accurate goodness of fit tests to see how accurate the $\chi^2$-test is in our scenario.

Since we are fixing the number of photons seen under both hypotheses to be equal, the distribution of photons is expected to have a multinomial distribution, with probabilities of observing a photon in each bin predicted by the different hypotheses.  We assume that we observe $o_i$ photons in $m$ bins consistent (up to rounding to the nearest integer) with the Standard Model hypothesis $H_\text{SM}$ (so $o_i = \text{Round}(N_\gamma p_i^{H_\text{SM}})$ where $p_i^{H_\text{SM}}$ is the probability of receiving a photon in the $i$-th bin).  We will call this Obs.  Then, given a BSM hypothesis $H_\text{BSM}$, which constitutes the probabilities of a photon landing in each bin under the BSM hypothesis, $p_i^{H_\text{BSM}}$, the expected number of photons in each bin is $e_i = N_\gamma p_i^{H_\text{BSM}}$.

We use a $p$-value to determine whether the observed photon spectrum (or something more extreme) is likely to come from chance alone. Finding the exact $p$-value of an observation is computationally expensive. We use the efficient \texttt{ExactMultinom} R package~\cite{Resin03042023}, which can quickly compute the $p$-value for less than around 1000 photons and fewer than around 8 bins.  Since in this example with decadal binning there are 10 bins and 1055 photons, we drop the longest two bins ($\tau \geq 10^{2}$\,s) to allow for comparison to the exact $p$-value.  The total number of photons is then $N_\gamma = 672$ and the distributions of observed and expected (BSM) events are
\begin{align}
\label{eq:example-data}
    o 
    &=
    \{1, 3, 9, 23, 54, 113, 197, 272\} \,,
    \\
    e
    &=
    \{0.8, 2.3, 6.2, 15.7, 40.5, 106.9, 207.6, 292.1\}
    \,.
\end{align}
For these observed and expected photon counts we find a $p$-value of $p_\text{exact} = 0.071$.

For more total photons or for more bins, this method takes a long time.  In these cases, the $p$-value can be estimated by a variety of methods.  One method is based on the log-likelihood ratio.  The test statistic is given by
\begin{align}
\label{eq:test-statistic}
    q_\text{obs}
    &=
    -2\log\left(\frac{L(\text{Obs}|H_\text{BSM})}{L(\text{Obs}|H_\text{SM})}\right)
    \,.
\end{align}
Note that when setting a sensitivity, $H_\text{BSM}$ is the null hypothesis and Obs is assumed to be equal to the SM expectation (up to rounding).  This is often called the Asimov data set.  In the case of a multinomial distribution, the likelihood $L$ is
\begin{align}
    L(\text{Obs}|H)
    &=
    N_\gamma!
    \prod_{i=1}^m 
    \frac{(p^H_i)^{o_i}}{o_i!}
    \,,
\end{align}
with
\begin{align}
    \sum_{i=1}^m o_i = N_\gamma
    \,.
\end{align}
Note that in the test statistic, \cref{eq:test-statistic}, the factors of $N_\gamma!$ and $o_i!$ cancel out, so that
\begin{align}
    q_\text{obs}
    &=
    -2\sum_{i=1}^m
    o_i \log\left(
    \frac{p_i^{H_\text{BSM}}}{p_i^{H_\text{SM}}}
    \right)
    =
    -2\sum_{i=1}^m
    o_i \log\left(
    \frac{e_i}{o_i}
    \right)
    \,.
\end{align}

\begin{figure}
    \centering
    \includegraphics[width=0.5\textwidth]{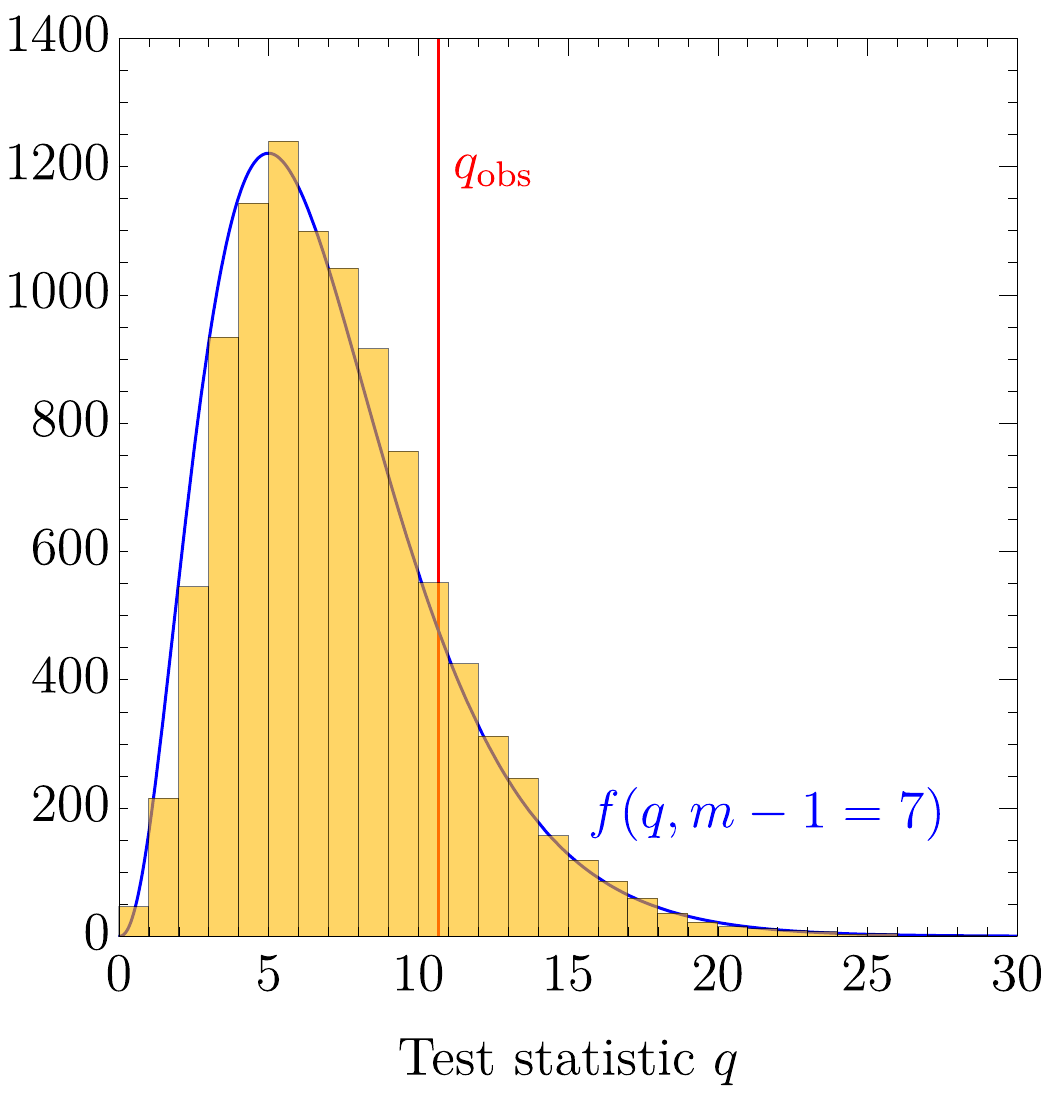}
    
    \caption{The distribution of the test statistic determined by Monte Carlo, the value of $q_\text{obs}$ for our example (vertical red line) and the $\chi^2$-distribution with $m-1 = 7$ degrees of freedom (in blue).
    }
    \label{fig:test-statistic}
\end{figure}

Since there will be randomness associated with the number of photons received, the distribution of the test statistic can be found through a Monte Carlo method, by drawing from the $H_\text{BSM}$ distribution many times and computing the value of the test statistic, to find the expected distribution of the test statistic from chance alone.  This distribution is shown in \cref{fig:test-statistic}.  The $p$-value can then be estimated by counting the number of times the test statistic was larger than or equal to $q_\text{obs}$ and dividing by the number of Monte Carlo samples.  Using this method we find $p_\text{MC} = 0.164$.

We see in \cref{fig:test-statistic} that the test statistic approximately follows a $\chi^2$-distribution with $m-1 = 7$ degrees of freedom.  This result leads to the G-test, where instead of taking Monte Carlo samples we simply compute
\begin{align}
    p_\text{G} 
    =
    \int_{q_\text{obs}}^\infty 
    f(x,m-1)
    dx
    \,,
\end{align}
where $f(x,m-1)$ is the probability density function of the $\chi^2$-distribution with $m-1=7$ degrees of freedom.  Using the G-test we find $p_\text{G} = 0.149$.

If the observed values $o_i$ are small fluctuations from the expected values $e_i$, $o_i = e_i + \delta_i$, and the sum of fluctuations is zero, $\sum_i \delta_i = 0$ (so there is the same number of total observed and expected photons), then
\begin{align}
    q_\text{obs}
    &=
    -2
    \sum_i 
    (e_i+\delta_i)\log\left(
        \frac{e_i}{e_i + \delta_i}
    \right)
    \\
    &=
    2
    \sum_i (e_i+\delta_i)
    \left(\frac{\delta_i}{e_i} - \frac{1}{2}\left(\frac{\delta_i}{e_i}\right)^2
    +\mathcal{O}\left(\left(\frac{\delta_i}{e_i}\right)^3\right)
    \right)
    \\
    &=
    \sum_i
    \frac{\delta_i^2}{e_i} +(e_i+\delta_i)\mathcal{O}\left(\left(\frac{\delta_i}{e_i}\right)^3\right)
    \\
    &\approx
    \sum_i \frac{(o_i-e_i)^2}{e_i}
    \,,
\end{align}
where we have used $\sum_i \delta_i = 0$.  As long as the fluctuations are small compared to the expected number of photons in all bins, the leading order term is a good approximation to the test statistic.  We can now see that the $\chi^2$-test is an approximation to the G-test.

To use the $\chi^2$-test we first compute the $\chi^2$-statistic,
\begin{align}
    \label{eq:chi-squared}
    \chi^2 = \sum_{i=1}^m \frac{(e_i-o_i)^2}{e_i}
    \,.
\end{align}
For a multinomial distribution, the $\chi^2$-statistic is expected to be $\chi^2$ distributed with $m-1$ degrees of freedom~\cite{ParticleDataGroup:2024cfk}.  The $p$-value is then given by
\begin{align}
    p_{\chi^2} = 
    \int_{\chi^2}^\infty
    f(x,m-1)
    dx
    \,.
\end{align}
For the distributions shown in \cref{fig:decadal-binning} we find $p_{\chi^2} = 0.108$.

We see that for this choice of expected and observed data, there is a reasonable difference in $p$-value between the exact ($p_\text{exact} = 0.071$),  Monte Carlo ($p_\text{MC} = 0.164$), G-test ($p_\text{G} = 0.149$) and $\chi^2$ ($p_{\chi^2} = 0.108$) methods.  The likely reason for this is that there are bins with few photons so the errors cannot be assumed to be Gaussian distributed.

\begin{figure}[h]
    \centering
    \includegraphics[width=0.475\textwidth]{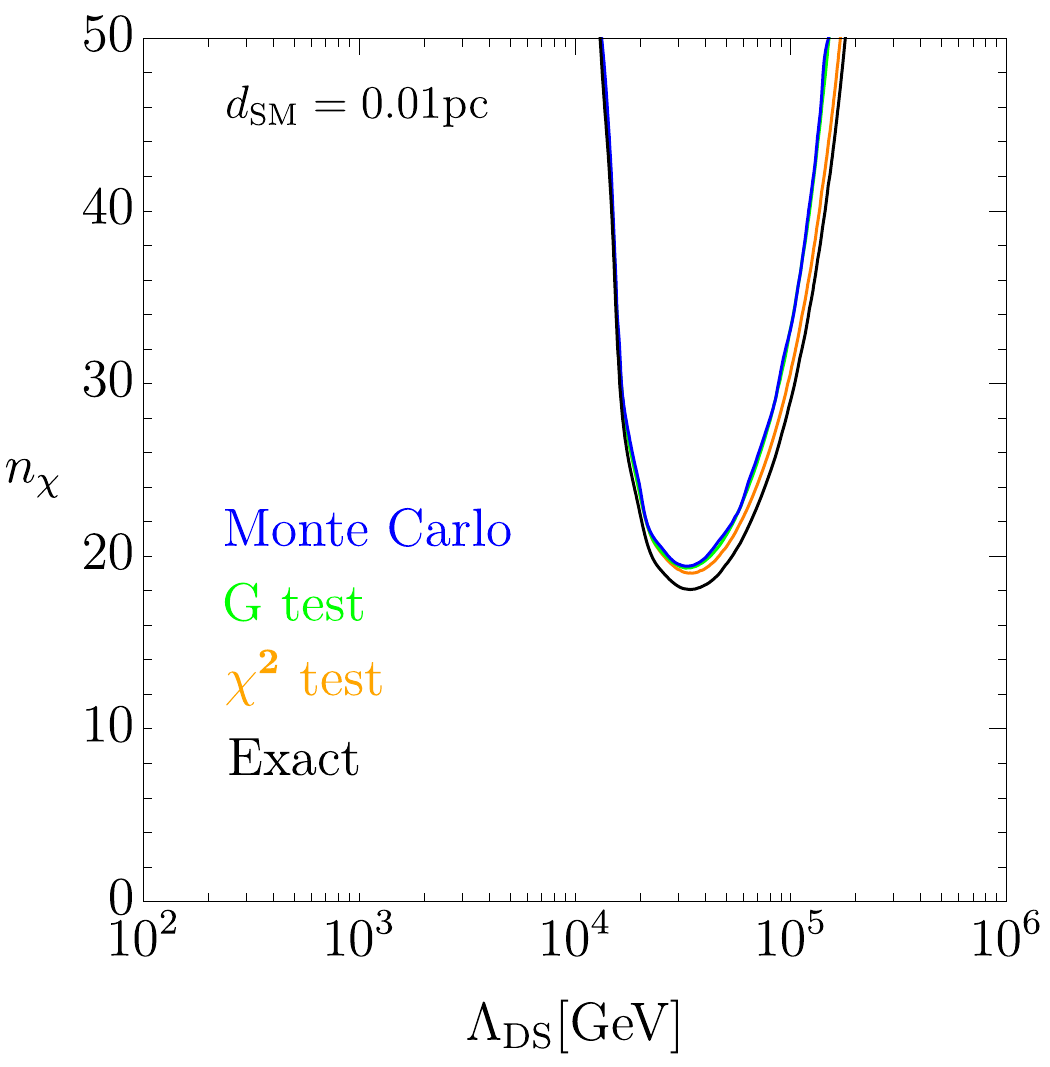}
    \includegraphics[width=0.475\textwidth]{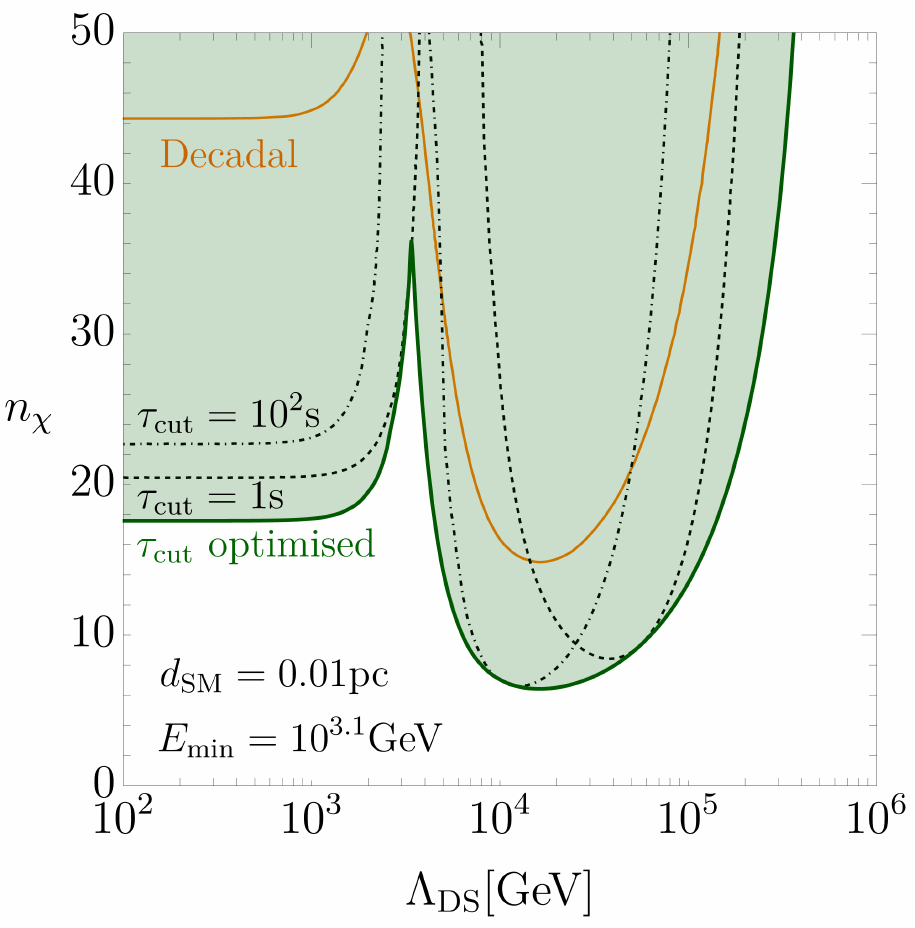}
    \caption{Left: Contours of $p=0.05$ for the exact, Monte Carlo, G- and $\chi^2$-tests for an explosion  at a distance of 0.01\,pc under the SM seen in the central portion of HAWC using decadal binning. We drop the longest two bins shown in \cref{fig:decadal-binning}. Right: Contour when the example data is combined into just two bins, following the prescription described in the text, compared to the contour obtained from the decadal binning and considering all 10 bins.}
    \label{fig:statistical-test-comparison}
\end{figure}

In \cref{fig:statistical-test-comparison} (left) we show the contour lines for $p=0.05$ (that is, where the DS model can be excluded at 95\% confidence level), determined using the four methods discussed above on the BSM parameter space. We see that the approximate methods of computing the $p$-value can introduce a small but non-negligible error in determining the parameter space probed by an observation.  Note that in this comparison we drop the longest two time bins.  We will see that when all the bins are kept there is improved sensitivity to new physics at lower energy scales (e.g., see \cref{fig:statistical-test-comparison} (right, orange curve) where we show the $\chi^2$ contour when considering all 10 bins).

\subsubsection{Comparison of Different Binning Strategies}

\begin{figure}[h]
    \centering
    \includegraphics[width=0.475\textwidth]{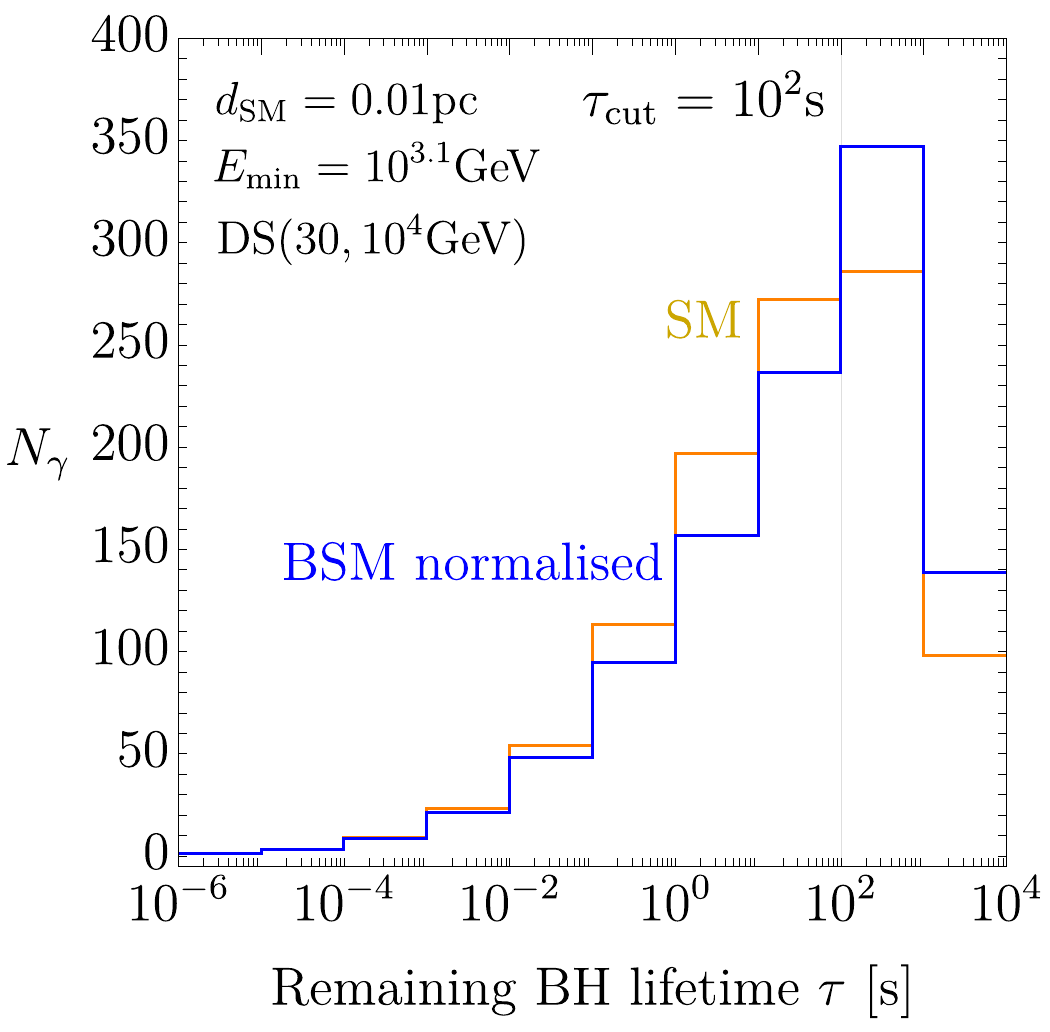}
    \includegraphics[width=0.475\textwidth]{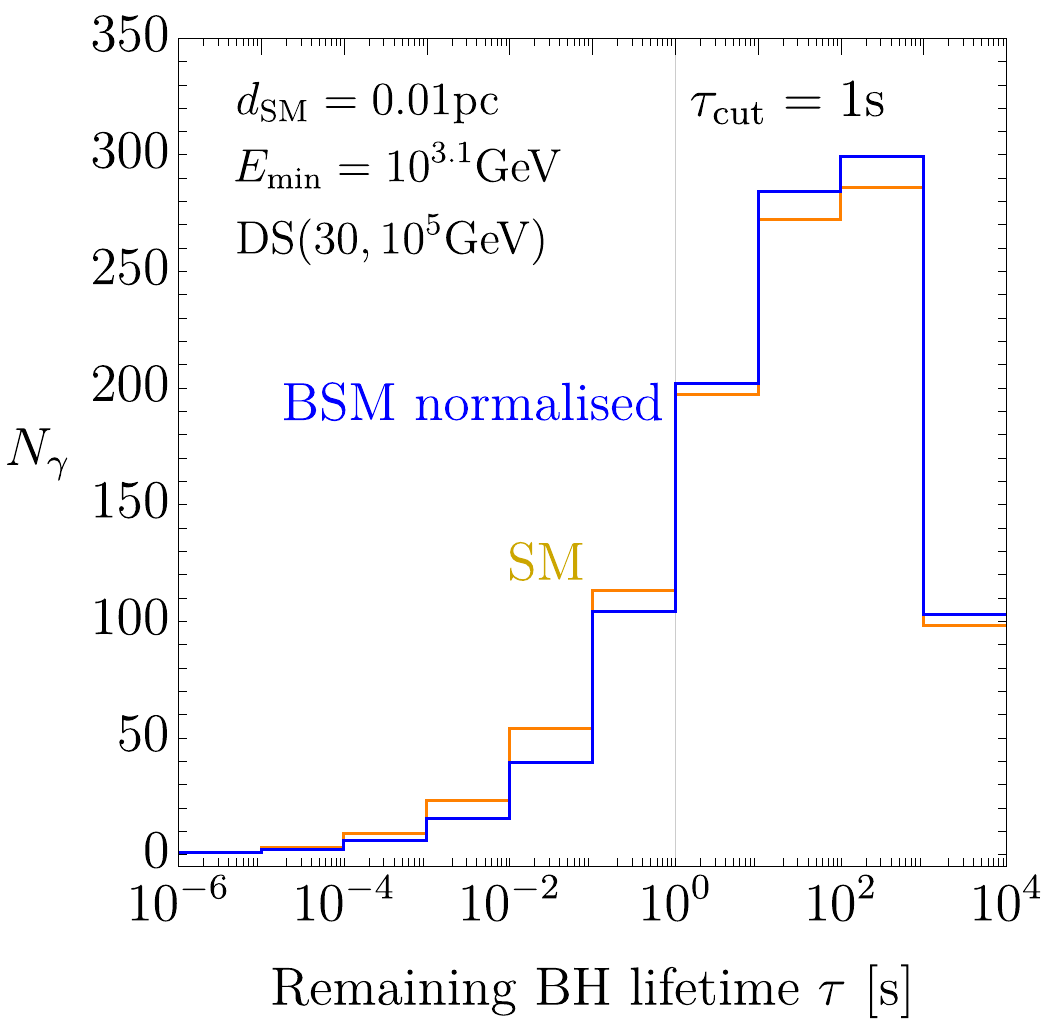}

    \caption{The signal from an exploding black hole seen in the central portion of HAWC with decadal binning, at a distance of 0.01\,pc in the SM scenario (orange) and after normalising the signal in the BSM scenario (blue).  On the left we show DS(30, $10^4$\,GeV) while the right shows DS(30, $10^5$\,GeV). The vertical gray line shows the position of the optimal $\tau_\text{cut}$.}
    \label{fig:mass-scale}
\end{figure}

We then explored a variety of methods of combining bins.  We found we could obtain the best sensitivity when the photons were binned into just two bins.  This has the side benefit that there are then enough photons in each bin that the $\chi^2$-method for computing the $p$-value gives a good approximation to the exact method.  In \cref{fig:statistical-test-comparison} (right) we compare the contours of $p = 0.05$ using the $\chi^2$-test statistic when the data given in \cref{fig:decadal-binning} is collected into just two bins.  We show the results obtained when we collect photon counts before and after $\tau_\text{cut} = 1$\,s (dashed), where the binned observed photon counts are
\begin{align}
    o 
    &=
    \{203, 852\}
    \,,
\end{align}
and the binned expected counts vary over the parameter space but at DS(30, $10^5$\,GeV) are
\begin{align}
    e
    &=
    \{167.7, 887.3\}
    \,.
\end{align}
We also show the result using $\tau_\text{cut} = 10^2$\,s (dash-dotted), where the binned observed and expected counts at DS(30, $10^5$\,GeV) are
\begin{align}
    o 
    &=
    \{672, 383\} \,,
    \\
    e
    &=
    \{654.3, 400.7\}
    \,.
\end{align}
We see that different choices of the bin separator can improve the sensitivity in some regions of parameter space, while making it worse in others.  We can understand this by looking at the normalised expected photon counts in decadal binning for BSM models with two different mass scales, e.g., DS(30, $10^4$\,GeV) and DS(30, $10^5$\,GeV), as shown in \cref{fig:mass-scale}.  For DS(30, $10^4$\,GeV), \cref{fig:mass-scale} (left), the new degrees of freedom become active around $10^3$\,s before the end of the explosion.  After normalisation there is then an excess of events in all bins with $\tau > 10^2$\,s and a deficit in all bins with $\tau < 10^2$\,s. We will get the best statistical sensitivity when we combine all bins with an excess into one bin, and all bins with a deficit into a second bin.   For DS(30, $10^5$\,GeV), \cref{fig:mass-scale} (right), there is an excess of events in all bins with $\tau > 1$\,s and a deficit in all bins with $\tau < 1$\,s.  As the BSM parameter point goes to higher mass scales, the new degrees of freedom turn on later and the optimal time to divide the binning shifts to shorter times before the end of the explosion.  We find that the optimal time to divide the two bins, $\tau_\text{cut}^\text{opt}$, depends on the point in BSM parameter space, the choice of $E_\text{min}$ (and therefore $\tau_\text{max}$) and the observatory we consider.  We therefore perform an optimisation procedure where, whenever we find a sensitivity, we vary $\tau_\text{cut}$ between $\tau_\text{min}$ and $\tau_\text{max}$ until we maximise the $\chi^2$-statistic found,
\begin{align}
    \chi^2(\tau_\text{cut}^\text{opt})
    &=
    \max_{\tau_\text{cut}\in[\tau_\text{min},\tau_\text{max}]}
    \chi^2(\tau_\text{cut})
    \,,
\end{align}
where $\chi^2(\tau_\text{cut})$ is the $\chi^2$-statistic given in \cref{eq:chi-squared} for a two bin test, with bin edges $[\tau_\text{min}, \tau_\text{cut}]$ and $[\tau_\text{cut}, \tau_\text{max}]$.  The result of this is given in the green region in \cref{fig:statistical-test-comparison} (right), where we choose $E_\text{min} = 10^{3.1}$\,GeV.  We see that it can be thought of as the envelope of the regions excluded when different $\tau_\text{cut}$'s are chosen.

Finally, in the following we also approximately incorporate systematic errors.  Since we are now using a two bin $\chi^2$-test we can simply use
\begin{align}
    \chi^2 = \sum_{i=1}^2 \frac{(e_i-o_i)^2}{e_i + (\sigma_\text{sys} e_i)^2}
    \,,
\end{align}
where we have assumed the errors to be uncorrelated and Gaussian, with a standard deviation $\sigma_\text{sys}$.  Rather than attempting to model the different experimental systematic errors in detail, we simply want to avoid implausible sensitivity when the photon counts are high so we take $\sigma_\text{sys} = 10\%$, which is a rough estimate of the experimental sensitivity of these gamma ray observatories.  E.g., in Ref.~\cite{HAWC:2019wla} HAWC's largest  systematic error was $11.76\%$.

\section{Results}
\label{sec:Results}

Now that we have discussed the current and future experiments we will consider, the signal and background for an exploding black hole and the optimal analysis strategy at a telescope with relatively poor energy resolution, we can compare choices of $E_\text{min}$, the different explosion scenarios depicted in \cref{fig:SkyMap} and observation in different gamma ray observatories.

\subsection{HAWC: Scenario I}
\label{sec:ScenarioI}

We first consider one of the most optimistic scenarios for HAWC: a black hole explodes within, and spends as much time as possible in, HAWC's $\theta_1$ band where the effective area is greatest across almost all energies,  see scenario I and the dark green tail in \cref{fig:SkyMap}.  This scenario has previously been considered in Ref.~\cite{Baker:2021btk} where conservative bounds of $E_\text{min}=10^4$\,GeV and $\tau_\text{max}=10^3$\,s were used to ensure a background free observation (in that work the background rates were set conservatively high).\footnote{Note that the dark sectors appearing in Ref.~\cite{Baker:2021btk} are labelled with number of dark copies of the SM, $N$, rather than the number of dark Dirac fermions, $n_\chi$, as used for this work.} This $\tau_\text{max}$ is small enough that the rotation of the Earth was neglected.

As discussed in \cref{sec:BGRates}, setting $N_\text{bg}^\gamma=1$ and choosing $E_\text{min}$ determines a maximum integration time $\tau_\text{max}$, see \cref{fig:MaxIntegrationTime}. Since the photon spectrum is enhanced at lower energies, \cref{fig:bh-emission} (left), a lower $E_\text{min}$ seems advantageous, but this comes with a shorter $\tau_\text{max}$, see \cref{fig:MaxIntegrationTime}, which reduces the total number of signal photons observed.  We consider the range of $E_\text{min}\geq 10^{3.1}$\,GeV (below this energy the cosmic ray effective area is not provided, and the gamma ray effective area drops quickly, \cref{fig:effective-areas}). We will see that in much of the parameter space HAWC is most sensitive when $E_\text{min}\sim 10^{3.4}\,\text{GeV} \sim 3\times10^3$\,GeV.

\begin{figure}
    \centering
    \begin{tabular}{cc}
    \includegraphics[width=0.45\textwidth]{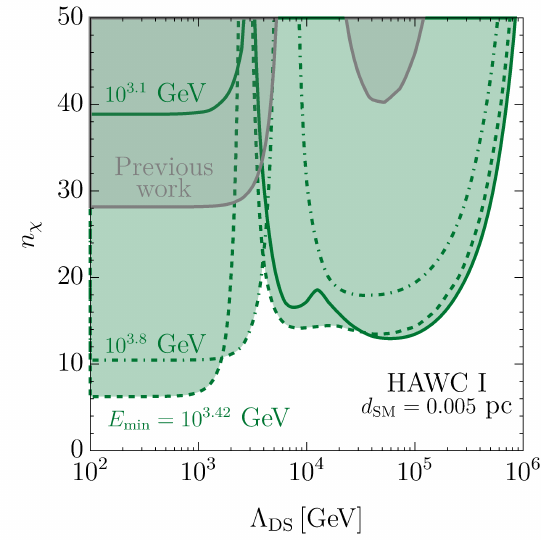}
    &
    \includegraphics[width=0.45\textwidth]{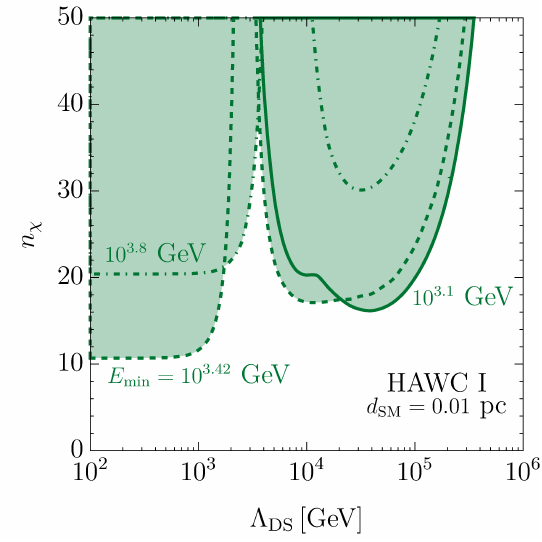}
    
    \end{tabular}
    \caption{The dark sector parameter space that can be probed by HAWC using different $E_\text{min}$ cuts under scenario I, assuming the number of photons received under the SM hypothesis at 0.005\,pc (left) and 0.01\,pc (right).  The $E_\text{min} = 10^{3.1}$\,GeV curve corresponds to $\tau_\text{max} = 2346$\,s, $E_\text{min} = 10^{3.42}$\,GeV to $\tau_\text{max} = 18\,322$\,s and $E_\text{min} = 10^{3.8}$\,GeV again to 18\,322\,s.  The "Previous work" reproduces the analysis from \cite{Baker:2021btk} but with updated instrument response functions.}
    \label{fig:HAWC_Method_Comparison_H1}
\end{figure}

For $E_\text{min}\sim 3\times10^3$\,GeV, $\tau_\text{max}$ at HAWC is of order $10^4$\,s. Over these timescales the proper motion of the Exploding Black Hole (EBH) can be ignored since the characteristic angular motion is $\sim1\degree$ per year for an EBH at $0.01$\,pc~\cite{Boluna:2023jlo}. However, we must account for the rotation of the Earth. As mentioned in \cref{sec:EBHSignalPhotons} the HAWC effective area depends on the angle of the EBH away from the zenith of HAWC, $\theta$.  As the Earth rotates throughout the observation, the EBH may cross between $\theta$ bands. Since HAWC is only $\sim10\degree$ North of the equator, to a good approximation the optimal path (scenario I) corresponds to an EBH that passes directly above HAWC.  If $E_\text{min}$ is large enough that $\tau_\text{max}\gtrsim 1.32\times10^4$\,s then the EBH will spend some observation time in the $\theta_2$ band.  If $\tau_\text{max}\gtrsim 1.60\times10^4$\,s then it will spend some time in the $\theta_3$ band.  At $\tau_\text{max}\approx 2\times10^4$\,s the EBH goes out of the HAWC field of view, so this sets an upper limit on the value for $\tau_\text{max}$, see \cref{sec:AppendixFOVGeometry}.\footnote{Around 24 hours earlier the black hole would also have been in HAWC's field of view, but at that time its temperature would be around $200$\,GeV and it would not produce photons that could be observed by HAWC.}  Note that one background event in the maximum observation time corresponds to $E_\text{min} = 10^{3.42}$\,GeV.

In \cref{fig:HAWC_Method_Comparison_H1} we show the 95\% C.L. ($p_{\chi^2} = 0.05$) exclusion regions for an EBH observed by HAWC in observation scenario I, using a two bin $\chi^2$-test with an optimised $\tau_\text{cut}$ and with $\sigma_\text{sys} = 10$\,\%. The distance to the exploding black hole in the SM scenario is $0.005$\,pc in the left panel and $0.01$\,pc in the right panel. The grey exclusion region shows the analysis method of Ref.~\cite{Baker:2021btk}, which uses decadal binning and a $\chi^2$-test with $E_\text{min}=10^4$\,GeV, $\tau_\text{max}=10^3$\,s and $\sigma_\text{sys} = 10$\,\%, applied to our updated instrument response functions.  

In \cref{fig:HAWC_Method_Comparison_H1} (left) we see that the two bin method significantly outperforms the method used in Ref.~\cite{Baker:2021btk}, reaching down to around $n_\chi = 6$ compared to $n_\chi = 28$ with the old approach.  The shape of the region that can be probed is similar in all cases: at $\Lambda_\text{DS}\lesssim10^3$\,GeV the new degrees of freedom are active for the whole time the black hole is observed, so the search is independent of the mass scale, for $10^3\,\text{GeV} \lesssim \Lambda_\text{DS} \lesssim 10^4$\,GeV there is a loss of sensitivity since after normalisation the time series of photons looks very similar in the SM and BSM cases, for $10^4\,\text{GeV} \lesssim \Lambda_\text{DS} \lesssim 3\times10^5$\,GeV there is very good sensitivity as HAWC would see the new degrees of freedom activate, while sensitivity is lost for $3\times10^5\,\text{GeV} \lesssim \Lambda_\text{DS}$ as HAWC receives very few photons from the final moments of the explosion when the PBH temperature reaches above this scale (for an explosion at 0.005\,pc).

Comparing the different choices of $E_\text{min}$ we see that there is a trade-off between reducing $E_\text{min}$ and reducing $\tau_\text{max}$.  The best sensitivity occurs for $E_\text{min} \sim 10^{3.42}\,\text{GeV} = 2.6\times10^3$\,GeV, which is the longest possible integration time with one background event, while sensitivity is generally reduced for smaller or larger $E_\text{min}$.  We see that while different choices of $E_\text{min}$ still feature a sensitivity loss in the region $10^3\,\text{GeV} \lesssim \Lambda_\text{DS} \lesssim 10^4$\,GeV, the exact behaviour is different in each case and the region can be filled in by taking two or more choices for $E_\text{min}$ (in this case $E_\text{min} \sim 10^{3.42}\,\text{GeV}$ and $E_\text{min} \sim 10^{3.8}\,\text{GeV}$).  Since we take the envelope of just three $E_\text{min}$'s there is a bump at the transition between them which could be filled in using intermediate $E_\text{min}$'s.  The smaller $E_\text{min} \sim 10^{3.1}\,\text{GeV}$ is slightly better than $E_\text{min} \sim 10^{3.42}\,\text{GeV}$ at mass scales $\Lambda_\text{DS} \gtrsim 3\times10^5$\,GeV but there is only marginal improvement.

In \cref{fig:HAWC_Method_Comparison_H1} (right) we show the same scenario but for the number of photons observed for an explosion occurring at a distance of $0.01$\,pc assuming the SM hypothesis.  Since the number of signal photons received goes as $1/d^2$, see \cref{{eq:EBHSignal}}, statistical errors are greater and less parameter space can be probed in this case.  While the method used in Ref.~\cite{Baker:2021btk} is not sensitive to $n_\text{Dirac} < 50$ at any mass scale, the 2-bin approach does not become dramatically weaker.  It is still sensitive to $n_\text{Dirac}\sim 10$ for $\Lambda_\text{DS}\lesssim10^3$\,GeV and to $n_\text{Dirac}\sim 15$ for $10^4 \lesssim \Lambda_\text{DS} \lesssim 10^5$\,GeV.  
The relative importance of the $E_\text{min}$ cuts remains unchanged. 

We also note, by comparing to \cref{fig:statistical-test-comparison} (right), that for $d_\text{SM}=0.01$\,pc and for $E_\text{min} \sim 10^{3.1}\,\text{GeV}$ systematic errors have a significant impact at $\Lambda_\text{DS} \lesssim 10^3$\,GeV, reducing the $n_\chi$ probed from 18 to more than 50, and a smaller but still significant impact at $10^4\,\text{GeV} \lesssim \Lambda_\text{DS} \lesssim 10^5$\,GeV, reducing the $n_\chi$ probed from 6 to 16.  For larger distances, the overall error will get a larger contribution from statistical errors and the impact of systematic errors will reduce.

\subsection{HAWC: Scenarios II and III}
\label{sec:ScenarioIIandIII}

In the previous section we considered the optimistic scenario in which an EBH is observed having spent as much time in the optimal HAWC zenith band as allowed by the rotation of the Earth and geometry of the field of view. However, an EBH is equally likely to explode at any point on the celestial sphere. We consider two further scenarios for an EBH observation at HAWC. Scenario II is less optimistic than scenario I, with the EBH exploding in the $\theta_3$ band but having spent some observation time in the more sensitive $\theta_2$ and (depending on $E_\text{min}$) $\theta_1$ bands. Scenario III is somewhat pessimistic, with the EBH exploding in the $\theta_3$ band having spent much of the time allowed by the background-free constraint entirely outside the HAWC field of view. The trajectories of the EBH are shown by the light green tails in \cref{fig:SkyMap}.

\begin{figure}
    \centering
    \begin{tabular}{cc}
    \includegraphics[width=0.45\textwidth]{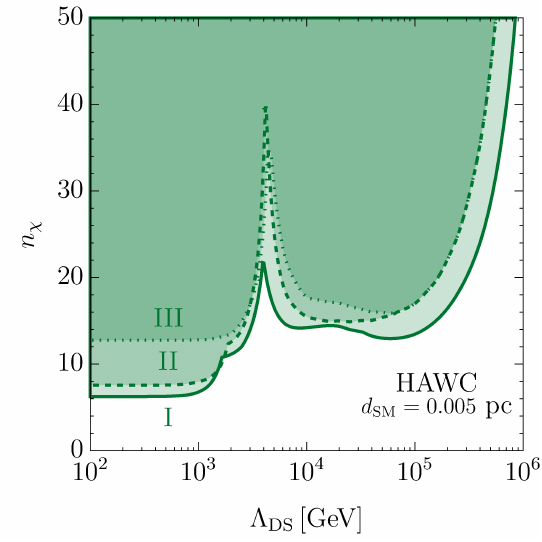} & 
    \includegraphics[width=0.45\textwidth]{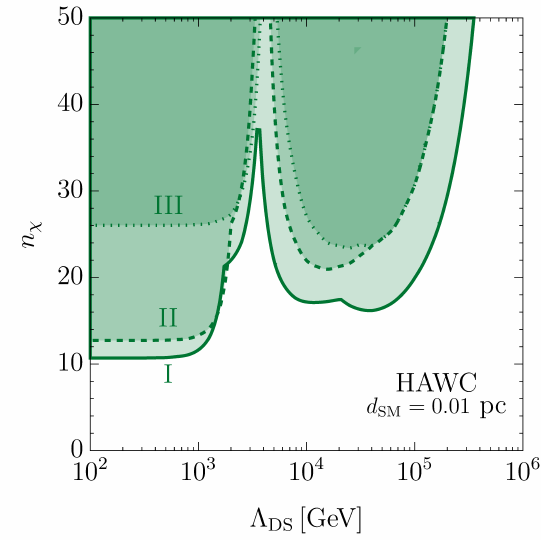}
    \end{tabular}
    \caption{
    Sensitivity of HAWC to dark sectors in various observational scenarios (see \cref{fig:SkyMap}).  See text for details.
    }
    \label{fig:HAWC_Scenarios_I_II_III_Comparison}
\end{figure}

\Cref{fig:HAWC_Scenarios_I_II_III_Comparison} compares the sensitivity of HAWC to a dark sector in scenarios I, II, and III. In these plots, each of the shaded regions is the envelope of the regions probed by the 2-bin method with $E_\text{min}=10^{3.1}$\,GeV, $E_\text{min}\approx10^{3.4}$\,GeV and $E_\text{min}=10^{3.8}$\,GeV.\footnote{Since the EBH spends a different amount of time in HAWC's field of view in the three different scenarios, the value of $E_\text{min}$ corresponding to exactly one background event depends on the scenario.  We find $E_\text{min} = 10^{3.42}$\,GeV, $E_\text{min} = 10^{3.50}$\,GeV and $E_\text{min} = 10^{3.39}$\,GeV for scenarios I, II and III, respectively.} On the left we show it for an explosion at a distance of $d_\text{SM} = 0.005$\,pc while on the right we show it for $d_\text{SM}=0.01$\,pc.  In scenario II the total number of photons observed is smaller than in scenario I due to the fact that the black hole is brightest at the end of its explosion and the smaller effective areas in the $\theta_{2,3}$ bands (shown in \cref{fig:effective-areas}). This means that the sensitivity in scenario II is slightly reduced in all of the parameter space compared to scenario I.   The situation is similar for black holes that explode at $d_\text{SM} = 0.005$\,pc (left) and $d_\text{SM} = 0.01$\,pc (right).

In scenario III, fewer photons are observed and the sensitivity is generally poorer than in scenarios I and II. However, even in this pessimistic scenario, where the explosion is entirely observed in the $\theta_3$ band, a significant region of parameter space can be probed. Dark sectors with $\Lambda_\text{DS} \gtrsim 10^{4}$\,GeV can be probed to a similar extent as in scenario II since the BSM degrees of freedom activate while the black hole is within the $\theta_3$ band in both scenarios. The sensitivity to lower dark sector scales is worse than in scenario I and II, since the black hole is not within HAWC's field of view when these dark sector states activate.  We note that in scenario III the region with $\Lambda_\text{DS} \lesssim 10^{3}$\,GeV is best probed using $E_\text{min}\approx10^{3.8}$\,GeV (rather than $E_\text{min}\approx10^{3.4}$\,GeV in scenarios I and II), so the optimal  $E_\text{min}$ in different regions of parameter space depends on the scenario under consideration.

\subsection{Current and Future Gamma Ray Telescopes: Scenarios IV, V, VI and VII}
\label{sec:ScenarioIV-VII}

In this section we compare benchmark observation scenarios at the various current and future gamma ray telescopes discussed in \cref{sec:Experiments}.  We recall that HAWC, LHAASO, SWGO and CTA North and South differ greatly in their effective areas as a function of the incoming gamma ray energy, \cref{fig:effective-areas}, and in their background rates, \cref{fig:backgrounds}.  

\begin{figure}
    \centering
    \begin{tabular}{cc}
       \includegraphics[width=0.45\textwidth]{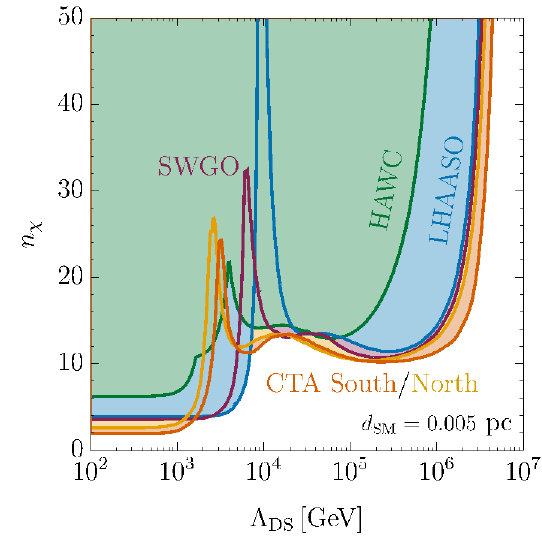}  & \includegraphics[width=0.45\textwidth]{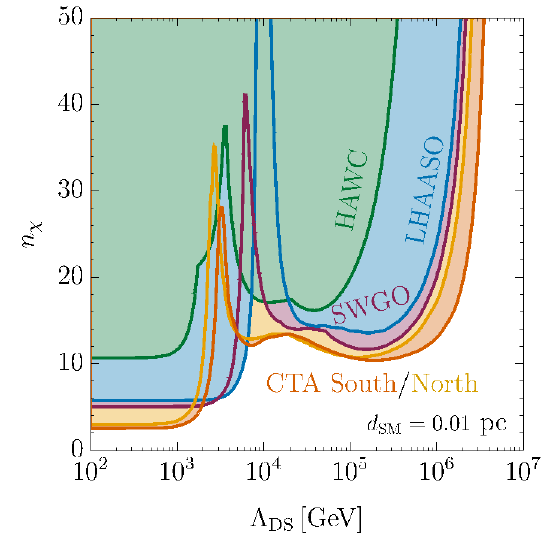}\\
       \includegraphics[width=0.45\textwidth]{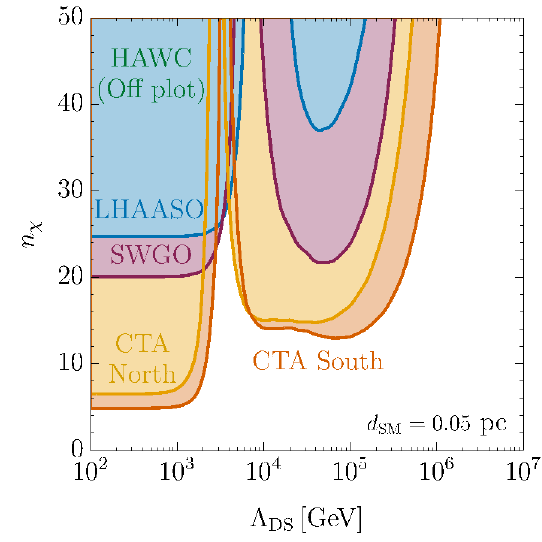}  & \includegraphics[width=0.45\textwidth]{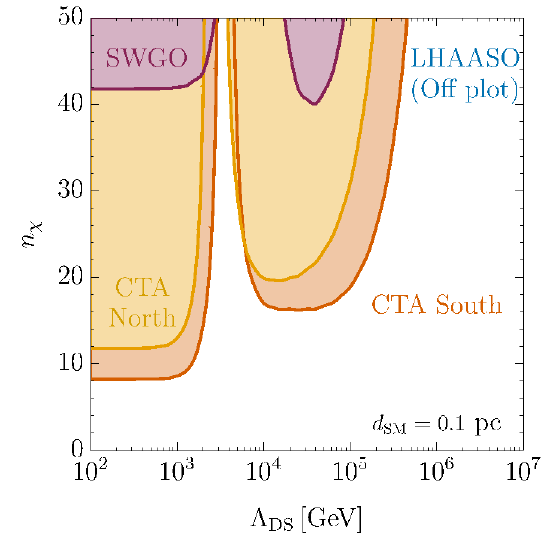}
    \end{tabular}
    \caption{
    Exclusion regions of the dark sector parameter space for explosions above HAWC, LHAASO, SWGO and CTA North and South, in the optimistic scenarios that the black holes spend as much time as needed in their fields of view. In the top left panel $d_\text{SM} = 0.005$\,pc while in the top right panel $d_\text{SM} = 0.01$\,pc. In the bottom left panel $d_\text{SM} = 0.05$\,pc and in the bottom right panel $d_\text{SM} = 0.1$\,pc. In each case we choose the $E_\text{min}$ which gives a good sensitivity over the parameter space. 
    The HAWC curve shows scenario I with the envelope of $E_\text{min} = 10^{3.1}$\,GeV, $10^{3.8}$\,GeV and $10^{3.42}$\,GeV as in \cref{fig:HAWC_Method_Comparison_H1}, LHASSO has $E_\text{min} = 10^4$\,GeV, SWGO has the envelope of $E_\text{min} = 10^{3.4}$\,GeV and $E_\text{min} = 10^{3.85}$\,GeV, CTA North has $E_\text{min} = 10^{3.45}$\,GeV while CTA South has $E_\text{min} = 10^{3.55}$\,GeV.
    }
    \label{fig:TelescopeComparison}
\end{figure}

In scenario IV, \cref{fig:SkyMap}, we assume that a black hole explodes above LHAASO.  We find the best sensitivity over the whole $\Lambda_\text{DS}$ range for $E_\text{min} \approx 10^4$\,GeV.  We show the expected sensitivity for an explosion at $d_\text{SM} = 0.005$\,pc in \cref{fig:TelescopeComparison} (upper left), assuming $\sigma_\text{sys} = 10\%$.  For LHAASO the background rates are high enough (\cref{fig:backgrounds} bottom right panel) that for $E_\text{min} = 10^4$\,GeV we expect one background event in $\tau_\text{max}\approx 700$\,s.  In this time the Earth rotates by only $\approx 3 \degree$. Therefore, as far as the angular dependence of the instrument response functions can be ignored, these results hold for an explosion almost anywhere in LHAASO's field of view.  We see that the final limits on $n_\chi$ are similar to HAWC in much of the parameter space. At low $\Lambda_\text{DS}$ LHAASO can reach down to $n_\chi \approx 4$ but loses sensitivity at $\Lambda_\text{DS} \approx 10^4$\,GeV, for similar reasons to HAWC's loss of sensitivity.  As in HAWC, this region could be probed by making different choices for $E_\text{min}$.  Around $\Lambda_\text{DS} \approx 3\times10^4$\,GeV the LHAASO and HAWC limits are both at $n_\chi \approx 12$.  However, at higher mass scales LHAASO has a significant advantage and can probe above $\Lambda_\text{DS} \approx 10^6$\,GeV.  Since very heavy dark sector particles activate closer to the end of the explosion when fewer photons are emitted, LHAASO's larger effective area helps it capture more of them and probe up to higher masses.  We also see the impact of the larger effective area for an explosion at further distances, in the other panels in \cref{fig:TelescopeComparison}.  At $d_\text{SM} = 0.01$\,pc (upper right panel) LHAASO is somewhat better than HAWC, while at $d_\text{SM} = 0.05$\,pc (lower left) LHAASO provides considerably stronger exclusions.  This is because at $d_\text{SM} = 0.005$\,pc both analyses were systematics limited but as the distance grows, the statistical errors dominate for HAWC sooner than for LHAASO.  At $d_\text{SM} = 0.1$\,pc (lower right) LHAASO no longer has sensitivity for $n_\chi < 50$.

In scenario V, \cref{fig:SkyMap}, we assume that a black hole explodes above SWGO.  We find good sensitivity over the whole parameter space by taking $E_\text{min} = 10^{3.4}$\,GeV and $E_\text{min} = 10^{3.85}$\,GeV, for which we find $\tau_\text{max}\approx 300$\,s and 2000\,s, respectively.  For $d_\text{SM} = 0.005$\,pc, \cref{fig:TelescopeComparison} (upper left), the exclusion limits are very similar to LHAASO's except the loss in sensitivity occurs at a slightly lower dark sector mass scale.  At further distances where the errors are statistics dominated, SWGO is more sensitive than LHAASO despite its similar effective area at high energies, \cref{fig:effective-areas}.  This is true at both large and small dark sector mass scales, and shows the importance of lower energy gamma rays (in this case around a TeV) in probing the BSM parameter space.  Again due to high background rates the Earth does not rotate significantly during the observation, so the results are valid for explosions occurring almost anywhere in SWGO's field of view.

In scenarios VI and VII we assume that a black hole explodes at the far West of CTA North and South's regions of coverage, respectively.  As discussed in \cref{sec:CTA_Overview}, CTA North and South will not have wide, static fields of view like HAWC or LHAASO but narrow fields of view that can be directed towards a point of interest on the sky.  They could therefore track the exploding black hole across a large portion of the sky (CTA North and South have a maximum zenith angle of $\theta=60\degree$~\cite{Gueta:2021vrf}) while minimising background with their good angular resolution, \cref{fig:backgrounds} (upper right).  We find the best sensitivities at $E_\text{min} = 10^{3.45}$\,GeV for CTA North and $E_\text{min} = 10^{3.55}$\,GeV for CTA South.  These correspond to $\tau_\text{max} \approx$ 33,000\,s (limited by the geometry rather than the background rate) and 31,000\,s, respectively.  Their large effective areas down to low energies is more important than their lack of sensitivity to gamma rays above $\approx 2\times 10^5$\,GeV, \cref{fig:effective-areas}, and CTA North and South provide the best sensitivity to black hole explosions around $d_\text{SM} = 0.1$\,pc, where the measurement for the other observatories is statistics limited.  The maximum distance that CTA South is sensitive to $n_\chi= 50$ is $d_\text{SM} = 0.58$\,pc at $\Lambda_\text{DS} \lesssim 10^3$\,GeV.

We can see from the sky map in \cref{fig:SkyMap} that, depending on where the explosion occurs, multiple experiments could observe the same event.  While this suggests that we could combine the results from different experiments and increase the projected sensitivity to the dark sector model, we find that the results from one telescope typically dominate over the others, and the sensitivity on the dark sector parameter space is only improved by a small amount.  However, in the event of an observation, it would be very important to compare and combine the data from different experiments, to get as full a picture of the event as possible.  We see from \cref{fig:SkyMap} that if the explosion occurs within HAWC's field of view, there will be the possibility to look for the pre-exploding black hole in multiple experiments (such as CTA North and LHAASO).  This could provide independent measurements of different phases of the explosion.  Furthermore, around 24 hours before the final explosion the black hole would have a temperature around $200$\,GeV and would produce photons that could be observed by CTA.  This means that if CTA observes a $200$\,GeV source, there is a chance that it could observe an explosion the following night.  We also note that black holes emit all stable particles, so there could also be independent measurements in neutrinos, positrons and anti-protons.

\section{Conclusions}
\label{sec:Conclusions}

In this work we consider the constraints that could be placed on the parameter space of a dark sector model consisting of $n_\chi$ fermions at a common mass scale $\Lambda_\text{DS}$ if an exploding Schwarzschild Black Hole were observed at HAWC, LHAASO, SWGO, CTA North or CTA South.  We take into account the grey body factors and radiation, hadronisation and/or decay of the primary particles using \texttt{BlackHawk}~\cite{Arbey:2019mbc,Arbey:2021mbl} and \texttt{HDMSpectra}~\cite{Bauer:2020jay}, respectively.  We use the latest instrument response functions from these key current and future gamma ray observatories to characterise the observed gamma ray signal and ensure a background free measurement.

In our analysis we do not make use of the energy of the incoming gamma rays, since this is reconstructed relatively poorly, but do make use of the gamma ray counts with time, since these experiments have excellent timing resolutions.  To ensure a background free observation we determine the relation between the minimum gamma ray energy and the maximum integration time.  We then explore various ways of binning the photons and consider different statistical tests, finding the best results when the photons are binned into two bins with a $\tau_\text{cut}$ which depends on the point in BSM parameter space, which we optimise on a point by point basis.  Although we could also optimise the minimum gamma ray energy on a point by point basis, we instead opt for taking the envelope of one, two or three different choices of $E_\text{min}$.

We find that relatively similar portions of BSM parameter space could be probed for an explosion almost anywhere in HAWC's field of view.  Comparing the different gamma ray telescopes, we find that they all perform similarly when limited by systematics (which we assume to be the same across the different telescopes) but that a large effective area and sensitivity to lower energy gamma rays are both important for probing the parameter space.  SWGO and LHAASO are expected to be able to probe more parameter space than HAWC, while CTA North and South will be able to probe the most.

Overall, we see that observation of an exploding black hole would be transformative, not just in providing experimental confirmation of primordial black holes and Hawking radiation, but also in providing definitive information about the particles present in nature.

\section*{Acknowledgements}

It is a pleasure to thank  Rafael Coelho Lopes de Sa for discussion of the statistical tests, and Kristi Engel and Mehr Un Nisa for discussion of HAWC specifications.  This work was partially supported by the Australian Government through the Australian Research Council Centre of Excellence for Dark Matter Particle Physics (CDM, CE200100008). We would also like to thank the IPPP at Durham University for their hospitality and partial support during this work.  This work was also performed in part at the Aspen Center for Physics, which is supported by National Science Foundation grant PHY-2210452. MJB and AT are also grateful to the Mainz Institute for Theoretical Physics (MITP) of the Cluster of Excellence PRISMA$^{+}$ (Project ID 390831469) for its hospitality and its partial support during the completion of this work.

\appendix

\section{Geometry of Fields of View}
\label{sec:AppendixFOVGeometry}

We consider several observation scenarios with observation times larger than $\mathcal{O}(10^4)$ seconds. In these cases the rotation of the Earth or, equivalently, the motion of the EBH on the celestial sphere cannot be ignored. An EBH will move along a line of constant latitude, $\phi_\text{EBH}$, from East to West. Since the Earth rotates with constant angular velocity, the amount of time an EBH spends in the field of view (FoV) of a given experiment is given by
\begin{align}
    \frac{\Delta t_\text{obs}}{1\,\text{day}}
    &=
    \frac{\Delta\lambda_\text{FoV}}{2\pi} \,,
\end{align}
where $\Delta\lambda_\text{FoV}$ is the longitudinal extent, i.e., the difference in longitude between the point at which the EBH enters the FoV and the point at which it either leaves the FoV due to diurnal rotation or the EBH totally evaporates. We will now discuss how to determine these two points. 

\begin{figure}[t]
    \centering
    \includegraphics[width=0.4\textwidth]{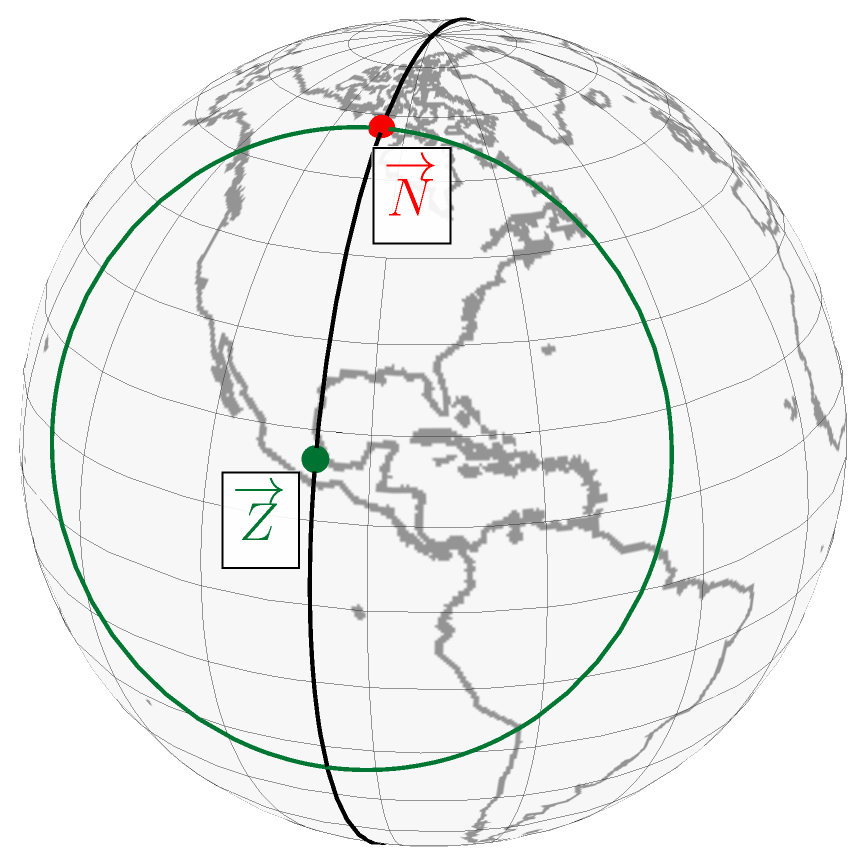} \qquad
    \includegraphics[width=0.3\textwidth]{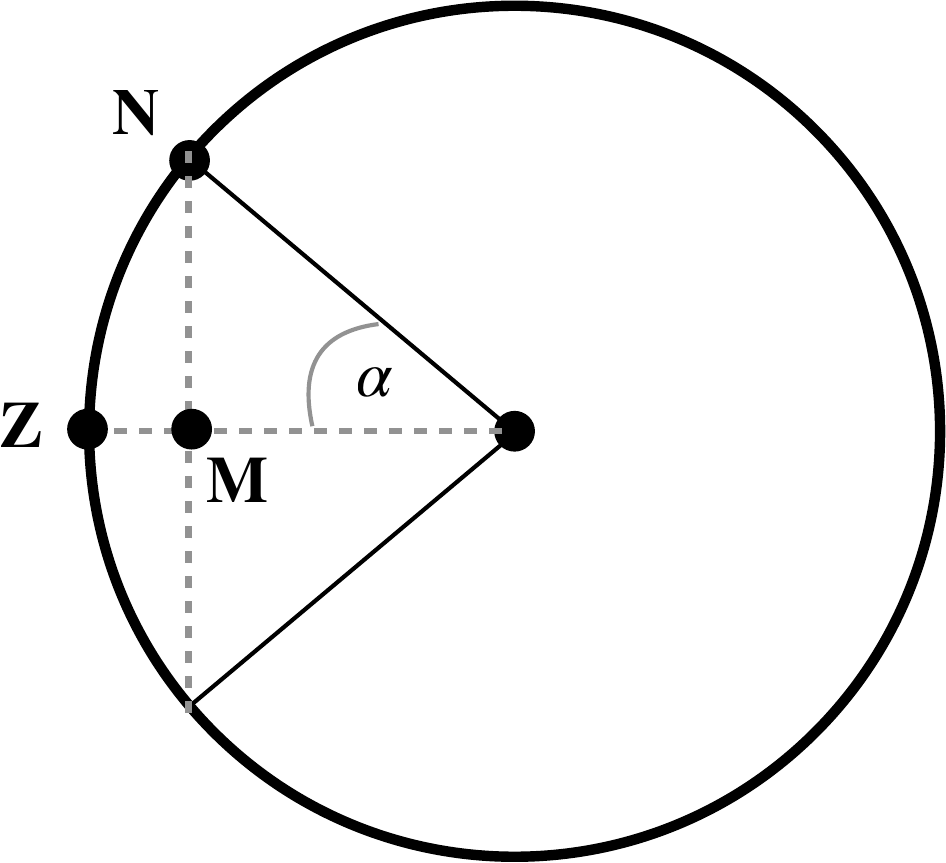}
    \caption{
    Geometry of the field of view cone boundary on the celestial sphere in 3d (left) and 2d (right).}
    \label{fig:SphericalGeo1}
\end{figure}

For any given experiment, we know the zenith point on the celestial sphere, and the angular radius $\alpha$ of the field of view. Treating the celestial sphere as a unit sphere we write the zenith unit vector as $\vec Z=(1, \frac{\pi}{2} - \phi_Z,\lambda_Z)$, where the vector is written in polar coordinates, $\phi_Z$ is the latitude and $\lambda_Z$ is the longitude. We now want to parameterise the circle that defines the intersection of the field of the view with the surface of the celestial sphere. To do this, we first identify any point on this circle and then rotate it around the axis defined by $\vec Z$ by an angle $\beta\in[0,2\pi)$. A convenient choice for the initial point is the point directly `northwards' of $\vec Z$ separated by an angle $\alpha$, $\vec N = (1, \frac{\pi}{2} - (\phi_Z + \alpha),\lambda_Z)$. To rotate $\vec N$ around the zenith, we define two orthonormal basis vectors, $\vec{e}_1$ and $\vec{e}_2$, that span the plane of the circle
\begin{equation}
\vec{e}_1 = \frac{\vec{N}-\vec{M}}{|\vec{N}-\vec{M}|}
,
\qquad
\vec{e}_2 = \vec{e}_1\times \vec{Z},
\end{equation}
where $\vec{M} = (\cos{\alpha}, \frac{\pi}{2} - \phi_Z,\lambda_Z)$ denotes the centre of the circle in this plane, see \cref{fig:SphericalGeo1} for a sketch of the geometry. With a radius of $\sin{\alpha}$, the boundary of the field of view can then be parameterised by 
\begin{equation}
\vec{V}(\beta) = 
\vec{M} + \sin{\alpha}\,(\cos{\beta}\,\vec{e}_1+\sin{\beta}\,\vec{e}_2),\qquad \beta\in[0,2\pi].
\end{equation}

An EBH will be observed at a certain latitude $\phi_\text{EBH}$. We can now numerically solve the equation $\phi_{\vec{V}(\beta)}=\phi_\text{EBH}$,  where $\phi_{\vec{V}(\beta)}$ is the latitudinal component of the vector $\vec V(\beta)$, to find the two solutions $\beta_{1,2}$ on the celestial sky where the EBH enters and exits the field of view (if it does not totally evaporate while in the FoV). The longitudinal extent is then given by $\Delta\lambda_\text{FoV} = |\lambda (\beta_1) - \lambda (\beta_2)|$ (or $2 \pi - \Delta\lambda_\text{FoV}$ if it is smaller than $\Delta\lambda_\text{FoV}$).

\bibliographystyle{JHEP}
\bibliography{refs}

\end{document}